\documentclass[fleqn,usenatbib,usedcolumn]{mnras}
\usepackage[british]{babel}
\usepackage{graphicx}
\usepackage{color,soul}
\usepackage{mathtools}
\usepackage{newtxtext}
\usepackage[slantedGreek]{newtxmath}
\usepackage[T1]{fontenc}
\usepackage{bm}
\usepackage{amsmath}
\usepackage{hyperref}
\usepackage{booktabs}
\usepackage[inline]{enumitem}
\usepackage{relsize}
\usepackage{todonotes}

\addto\extrasbritish{%
}

\let\mr\mathrm
\DeclarePairedDelimiter\abs{\lvert}{\rvert}%
\graphicspath{{Figures/}}
\newcommand{\HII}{\mbox{H\,{\sc ii}} }
\newcommand{\lan}{\langle }
\newcommand{\ran}{\rangle }
\newcommand{\sTheta}{\bm{\upTheta}} 
\newcommand{\sPsi}{\bm{\upPsi}}     
\newcommand{\sSigma}{\bm{\upSigma}} 
\newcommand{\sdelta}{\deltaup}      
\newcommand{\sDelta}{\Deltaup}      
\newcommand{\sLambda}{\upLambda}    

\begin{document}

\title[Joint estimation of the Epoch of Reionization power spectrum and foregrounds]{Joint estimation of the Epoch of Reionization power spectrum and foregrounds}
\author[Sims et al.]{Peter H. Sims$^1$\thanks{E-mail: peter\_sims1@brown.edu}, Jonathan C. Pober$^1$ \\
$^1$Department of Physics, Brown University, Providence, RI 02912, USA \\
}

\maketitle
\label{firstpage}
\begin{abstract}
The power spectrum of redshifted 21 cm emission brightness temperature fluctuations is a powerful probe of the Epoch of Reionization (EoR). However, bright foreground emission presents a significant impediment to its unbiased recovery from interferometric data. We build on the Bayesian power spectral estimation methodology introduced in \citet{2016MNRAS.462.3069S} and demonstrate that incorporating a priori knowledge of the spectral structure of foregrounds in the large spectral scale component of the data model enables significantly improved modelling of the foregrounds without increasing the model complexity. We explore two astrophysically motivated parametrisations of the large spectral scale model: \begin{enumerate*}\item a constant plus power law model of the form $q_{0}+q_{1}(\nu/\nu_{0})^{b_{1}}$ for two values of $b_{1}$: $b_{1} = <\beta>_\mathrm{GDSE}$ and $b_{1} = <\beta>_\mathrm{EGS}$, the mean spectral indices of the Galactic diffuse synchrotron emission and extragalactic source foreground emission, respectively, and \item a constant plus double power law model of the form $q_{0}+q_{1}(\nu/\nu_{0})^{b_{1}}+q_{2}(\nu/\nu_{0})^{b_{2}}$ with $b_{1} = <\beta>_\mathrm{GDSE}$ and $b_{2} = <\beta>_\mathrm{EGS}$. \end{enumerate*} We estimate the EoR power spectrum from simulated interferometric data consisting of an EoR signal, Galactic diffuse synchrotron emission, extragalactic sources and diffuse free-free emission from the Galaxy. We show that, by jointly estimating a model of the EoR signal with the constant plus double power law parametrisation of the large spectral scale model, unbiased estimates of the EoR power spectrum are recoverable on all spatial scales accessible in the data set, including on the large spatial scales that were found to be contaminated in earlier work.
\end{abstract}


\section{Introduction}
\label{Introduction}

The birth of the first stars and galaxies at Cosmic Dawn (CD) and the subsequent Epoch of Reionization (EoR), when these first luminous sources became abundant enough to drive a global phase change in the intergalactic medium (IGM), are among the least observed eras of cosmic history. The study of this era will enable us to constrain cosmological parameters (e.g. \citealt{2006ApJ...653..815M, 2008PhRvD..78b3529M, 2009MNRAS.394.1667F, 2016PhRvD..93d3013L}), probe directly the initial stages of structure formation, and characterise the properties of the first stars, proto-galaxies and accreting black holes (e.g. \citealt{2012MNRAS.424..762D, 2013MNRAS.431..621M, 2014MNRAS.439.3262M, 2015MNRAS.449.4246G}).

The redshifted 21 cm hyperfine line emission from the neutral hydrogen that pervades the IGM prior to the completion of reionization is a powerful probe of this period (see e.g. \citealt{2006PhR...433..181F, 2007RPPh...70..627B, 2010ARA&A..48..127M, 2012RPPh...75h6901P}). The intensity of the redshifted 21 cm emission can be measured as a differential brightness temperature between the 21 cm spin temperature and the brightness temperature of the radio background. Experiments to measure both the evolution of the sky-averaged `global' redshifted 21 cm signal and fluctuations in its intensity as a function of spatial scale are underway. 

The global signal, targeted principally by single-dipole experiments, traces the sky-averaged ionization history of the hydrogen IGM and constrains the timing of CD and the EoR. The first reported detection of the global 21 cm signal by the Experiment to Detect the Global Epoch of Reionization Signature (EDGES; \citealt{2018Natur.555...67B}) finds an absorption trough\footnote{However, see \citet{2018Natur.564E..32H} for concerns regarding the modelling of foregrounds in the analysis of the EDGES data which call into question the interpretation of results as an unambiguous detection of the cosmological 21-cm absorption signature.} centred at $78~\mathrm{MHz}$, with a width of $19~\mathrm{MHz}$. A high-redshift absorption trough is expected to result from Lyman-$\alpha$ photons from the first luminous sources coupling the 21 cm spin temperature to the kinetic temperature of the hydrogen gas, which has cooled through adiabatic expansion relative to the CMB, via the Wouthuysen-Field effect (\citealt{1952AJ.....57R..31W, 1958PIRE...46..240F, 1959ApJ...129..525F}). The 21 cm line has a rest-frame frequency of $1420~\mathrm{MHz}$. Expansion of the Universe redshifts the line to an observed frequency according to $\nu = 1420/(1+​z)~\mathrm{MHz}$, where $z$ is the redshift. Thus, the observed feature places CD at $z \lesssim 20$ (approximately $180~\mathrm{Myr}$ after the Big Bang). Observations of the Gunn-Peterson trough in high redshift quasar spectra (e.g. \citealt{2006AJ....132..117F}) place upper limits on the hydrogen neutral fraction in the IGM and imply that reionization is complete, or very nearly complete, by $z=6$. Measurements of the cosmic microwave background (CMB; e.g. \citealt{2016A&A...596A.108P}) find that the average redshift at which reionization occurs lies between $z = 7.8$ and 8.8 and, combined with additional measurements of the amplitude of the kinetic Sunyaev-Zeldovich from the higher-resolution Atacama Cosmology Telescope and South Pole Telescope experiments, constrain the duration of reionization to $\sDelta z < 2.8$. Measurements of  damping wing absorption (e.g. \citealt{2017MNRAS.466.4239G}) and the statistics of Lyman Break Galaxies (e.g. \citealt{2018ApJ...856....2M}) further constrain the ionization history of the hydrogen IGM and indicate a midpoint of reionization, characterised by a $50\%$ ionization fraction, at $z\sim7$.

Despite these exciting developments, the vast majority of the information encoded in the EoR signal remains untapped. Brightness temperature fluctuations in the redshifted 21 cm signal, when detected, will provide a multi-spatial scale probe of the ionisation, density and temperature state of the IGM during the EoR. Their measurement can be used to infer the spatial distribution and radiative properties of the sources driving reionization. The power spectrum of temperature fluctuations in the redshifted 21 cm signal is a high signal-to-noise statistic that encodes much of the information. As such, it is the initial target of experiments designed to detect fluctuations in the EoR signal. A number of interferometric experiments are aiming to detect the redshifted 21 cm power spectrum of the EoR. These include: the Giant Metrewave Radio Telescope (GMRT; \citealt{2013MNRAS.433..639P})\footnote{http://www.gmrt.ncra.tifr.res.in}, the LOw Frequency ARray (LOFAR; \citealt{2013A&A...556A...2V})\footnote{http://www.lofar.org/}, the Murchison Widefield Array (MWA; \citealt{2013PASA...30....7T})\footnote{http://www.mwatelescope.org/}, the Donald C. Backer Precision Array for Probing the Epoch of Reionization (PAPER; \citealt{2010AJ....139.1468P})\footnote{http://eor.berkeley.edu/}, the Hydrogen Epoch of Reionization Array (HERA; \citealt{2017PASP..129d5001D})\footnote{http://reionization.org/} and, in the near future, the Square Kilometre Array (SKA; \citealt{2013ExA....36..235M})\footnote{https://www.skatelescope.org/}. 

Amongst the most significant challenges faced by experiments aiming to detect the power spectrum of the EoR is the extraction of unbiased estimates of the signal in the presence of intense foreground emission. The total power in astrophysical emission at radio frequencies is dominated by Galactic diffuse synchrotron emission (GDSE) and synchrotron emission from extragalactic sources (EGS). In the sub-$200~\mathrm{MHz}$ frequency range of interest for detection of redshifted 21 cm emission from the Epoch of Reionization (EoR), this emission exceeds that from the EoR signal by up to five orders of magnitude in intensity.

Significant effort has been dedicated to investigating methods for recovering estimates of the EoR power spectrum in the presence of foregrounds. The proposed methods can be categorised as being drawn from one of two classes: \begin{enumerate} \item those seeking to avoid foreground contamination by taking advantage of the separation of spectrally smooth and rapidly fluctuating foreground and EoR signals, respectively, and estimating the power spectrum in a wedged shaped `EoR window' in $k_{\perp}$--$k_{\parallel}$-space\footnote{Here, $k$ is the Fourier conjugate variable to $r$, the comoving distance, and $k_{\perp}$ and $k_{\parallel}$ are the components of $k$ in the directions perpendicular and parallel to the line of sight, respectively.}, where contamination by smooth spectrum foregrounds is minimised (e.g. \citealt{2012ApJ...756..165P, 2013ApJ...768L..36P, 2016ApJ...833..242L}).
\item those seeking to recover estimates of the EoR power spectrum across the full range of spatial scales accessible in the data set, either by calculating the power spectrum of the residuals following subtraction of a foreground model (e.g. \citealt{2006ApJ...648..767M, 2009ApJ...695..183B, 2011PhRvD..83j3006L, 2012MNRAS.423.2518C, 2013MNRAS.429..165C, 2015MNRAS.447.1973B, 2018MNRAS.478.3640M}) or by jointly estimating a model for the foregrounds and EoR signal (e.g. \citealt{2016MNRAS.462.3069S, 2019MNRAS.484.4152S}). \end{enumerate}

In \citet{2016MNRAS.462.3069S} (hereafter S16) it was demonstrated that the inclusion of an accurate instrumental model, in combination with a quadratic model for intrinsic spectral structure in the foregrounds on spectral scales in excess of the 8 MHz bandwidth used in the analysis, is sufficient to recover unbiased estimates of the EoR power spectrum on intermediate and small spatial scales ($\log_{10}(k[h\mathrm{Mpc^{-1}}]) > -0.80$), in the presence of foregrounds, from simulated observations with the Hydrogen Epoch of Reionization Array (HERA; \citealt{2017PASP..129d5001D}). This was shown both for cold regions lying out of the plane of the Galaxy and for a region of more intense Galactic emission overlaying the Galactic plane and resulting in more pervasive power spectral contamination. In both cases, it was found that, on larger spatial scales, GDSE with a spatially dependent spectral index distribution and, to a lesser extent, synchrotron emission from the diffuse sea of extragalactic sources below the confusion noise limit of the instrument are sufficiently intense to dominate the recovered power spectral estimates. 

In this paper, we demonstrate that the region of $k$-space intrinsically accessible for estimation of the EoR power spectrum, within our analysis framework, can be expanded by improving upon the generic assumption of foreground spectral smoothness, used previously, with explicit incorporation of a priori knowledge of the spectral structure of foregrounds in our large spectral scale model.
To that end, we explore two astrophysically motivated parametrisation of the large spectral scale model: \begin{enumerate*}\item a constant plus power law model of the form $q_{0}+q_{1}(\nu/\nu_{0})^{b_{1}}$ for two values of $b_{1}$: $b_{1} = <\beta>_\mathrm{GDSE}$ and $b_{1} = <\beta>_\mathrm{EGS}$, the mean spectral indices of the Galactic diffuse synchrotron emission and extragalactic source foreground emission, respectively, and \item a constant plus double power law model of the form $q_{0}+q_{1}(\nu/\nu_{0})^{b_{1}}+q_{2}(\nu/\nu_{0})^{b_{2}}$ with $b_{1} = <\beta>_\mathrm{GDSE}$ and $b_{2} = <\beta>_\mathrm{EGS}$.\end{enumerate*}

The remainder of this paper is organised as follows. In \autoref{PowerSpectralModel}, building on the work in S16 and \citet{2019MNRAS.484.4152S} (hereafter, S19), we summarise our Bayesian approach to power spectral estimation and describe the implementation of the improved large spectral scale model. In \autoref{SkyModels}, we describe our EoR signal, foreground simulations, and instrument model. In \autoref{Analysis} we analyse the EoR power spectral estimates recovered from our simulated EoR plus foreground data when using each of our large spectral scale models and for foreground intensities expected for a HERA-like choice of instrument model. We demonstrate the effectiveness of incorporating each of our improved large spectral scale models, within our Bayesian power spectral estimation framework, for recovery of unbiased estimates of the EoR power spectrum. We summarise our conclusions and discuss future work in \autoref{Conclusions}.

\section{Power spectral estimation}
\label{PowerSpectralModel}

The method we use to estimate the intrinsic power spectra of the EoR and foregrounds in this work relies on Bayesian inference and builds on the approach demonstrated in S16 and described in detail in S19. We refer the interested reader to S19 for a detailed description of the methodology and to S16 for its application to recovery of the intrinsic power spectrum of the EoR from simulated observations with HERA, when using a quadratic large spectral scale model. In this section, we first outline key components of the data and power spectral models, and in \autoref{AMLSCM} we introduce the astrophysically motivated parametrisation of the large spectral scale component of our data model adopted in this paper.

\subsection{Bayesian inference}
\label{BayesianInference}

Our method for analysing the intrinsic power spectra of the EoR is built upon the principles of Bayesian inference, which provides a consistent approach to the estimation of a set of parameters, $\sTheta$, from a model, $M$, given a set of data $\bm{D}$. Bayes' theorem states that:
\begin{equation}
\label{Eq:BayesEqn}
\mathrm{Pr}(\sTheta\vert\bm{D},M) = \dfrac{\mathrm{Pr}(\bm{D}\vert\sTheta,M)\ \mathrm{Pr}(\sTheta\vert M)}{\mathrm{\mathrm{Pr}}(\bm{D}\vert M)} = \dfrac{\mathcal{L}(\sTheta)\pi(\sTheta)}{\mathcal{Z}}, 
\end{equation}
where $\mathrm{Pr}(\sTheta\vert\bm{D},M)$ is the posterior probability distribution of the parameters, $\mathrm{Pr}(\bm{D}\vert\sTheta,M) \equiv \mathcal{L}(\sTheta)$ is the likelihood, $\mathrm{Pr}(\sTheta\vert M) \equiv \pi(\sTheta)$ is the prior probability distribution of the parameters and $\mathrm{Pr}(\bm{D}\vert M)\equiv\mathcal{Z}$ is the Bayesian evidence. 

Since the evidence is independent of the parameters $\sTheta$, to make inferences regarding the model parameters we sample from the unnormalised posterior,
\begin{equation}
\label{Eq:UnnoramisedPosterior}
\mathrm{Pr}(\sTheta\vert\bm{D},M) \propto \mathcal{L}(\sTheta)\pi(\sTheta).
\end{equation}

\subsection{S16 data model}
\label{S16DataModel}

In S19, two approaches to recovering the power spectrum of the EoR are discussed. The preferred approach is dependent on the number of $k$-space amplitude parameters required to model the data. This number, in turn, is determined by the sampling rate (the sampling of the $uv$-plane and the channel width) and $k$-space volume (dependent on the range of baseline lengths, the bandwidth of the observation, and primary beam of the instrument) of the interferometric data set from which the power spectrum is to be estimated.
\begin{enumerate}
\item When a large number of signal coefficients are required to model the data, the computational expense associated with direct calculation of $\mathrm{Pr}(\bm{\varphi}|\mathbfit{d})$, with $\bm{\varphi}$ the intrinsic power spectrum of the signal and $\mathbfit{d}$ the data set from which we seek to estimate the power spectrum of the EoR, is significant. In this case, it is more computationally efficient to sample from the high dimension joint posterior probability density function, $\mathrm{Pr}(\bm{\varphi}, \mathbfit{a}, \mathbfit{q}|\mathbfit{d})$, of the intrinsic power spectrum and of a set of $k$-space amplitude parameters, $\mathbfit{a}$ and $\mathbfit{q}$, on well-sampled spatial scales and for large spectral scale fluctuations along the $\eta$-axis, respectively, with $\eta$ the Fourier dual to frequency $\nu$. Marginalisation to recover $\mathrm{Pr}(\bm{\varphi}|\mathbfit{d})$ can subsequently be performed numerically.
\item When the number of $k$-space amplitude parameters required to describe the signal is small ($\mathcal{O}(10000)$ or less), the coefficients can be marginalised over analytically and the power spectrum coefficients can be sampled from directly. In this case, we can sample from the far smaller parameter space of the intrinsic power spectrum of the signal alone and estimate the posterior probability density function for the power spectrum of the EoR, $\mathrm{Pr}(\bm{\varphi}|\mathbfit{d})$, directly from the data.
Evaluating $\mathrm{Pr}(\bm{\varphi}|\mathbfit{d})$ requires inverting a matrix associated with the analytic marginalisation. For a dense matrix inversion, this scales with the number of $k$-space amplitude parameters cubed.
\end{enumerate}

In this work, we take advantage of the significantly reduced number of parameters, and correspondingly improved computational efficiency, enabled by performing our analysis in the small $k$-cube regime, applicable to observations with HERA on sub-40 m baselines, and sample from $\mathrm{Pr}(\bm{\varphi}|\mathbfit{d})$ directly. However, because $\mathrm{Pr}(\bm{\varphi}|\mathbfit{d})$ is derived by marginalising over $\mathrm{Pr}(\bm{\varphi}, \mathbfit{a}, \mathbfit{q}|\mathbfit{d})$ analytically, in the remainder of this section we start by deriving the form of $\mathrm{Pr}(\bm{\varphi}, \mathbfit{a}, \mathbfit{q}|\mathbfit{d})$ given in S16. In \autoref{MarginalisationOvertheSignalCoefficients} we marginalise over the coefficients $\mathbfit{a}$ and $\mathbfit{q}$ to yield $\mathrm{Pr}(\bm{\varphi}|\mathbfit{d})$. Finally, in \autoref{AMLSCM} we describe the astrophysically motivated large spectral scale structure model, the analysis of which is the focus of this work.

S16 begin by defining a Gaussian likelihood function for the data model,
\begin{eqnarray}
\label{Eq:BasicVisLike}
\mathcal{L}(\mathbfit{a}, \mathbfit{q}) \propto \frac{1}{\sqrt{{\mathrm{det}(\mathbfss{N})}}} \exp\left[-\frac{1}{2}\left(\mathbfit{d} - \mathbfit{m}(\mathbfit{a}, \mathbfit{q})\right)^{\dagger}\mathbfss{N}^{-1}\left(\mathbfit{d} - \mathbfit{m}(\mathbfit{a}, \mathbfit{q})\right)\right] \ ,
\end{eqnarray}
where $\mathbfit{d} = \mathbfit{s} + \sdelta\mathbfit{n}$ is the data vector and is comprised of the signal $\mathbfit{s}$ and of noise $\sdelta\mathbfit{n}$. The signal (corrupted by noise) is, in principle, observed; in this paper, it is obtained from simulated image cubes through the two-dimensional discrete Fourier transform (DFT) to $uv$-space of each channel. The noise in the $uv$-domain is modelled as an uncorrelated Gaussian random field, with covariance matrix $\mathbfss{N}$. The elements of the covariance matrix are given by $\mathbfss{N}_{ij} = \left< n_in_j^*\right> = \delta_{ij}(\sigma_{i}^{2}+\alpha_{j}^{2})$. Here, $\left< ... \right>$ represents the expectation value, $\sigma_{j}$ is the RMS value of the noise in visibility element $j$ and $\alpha_j$ is a small-scale-structure model parameter which accounts for the high-frequency structure on scales smaller than the channel width, which manifests itself as an additional source of noise.

As in S16, here we want to make inferences regarding the $k$-space power spectrum of the signal from the $uv$-domain representation of our EoR and foreground simulations and construct our data model, $\mathbfit{m}(\mathbfit{a}, \mathbfit{q})$, via a matrix transformation, from a three-dimensional grid in $k$-space, $K_{m}(k_x, k_y, k_z)$, to their measurement-domain representation, $V_{m}({u, v, \nu_{i}})$. In \autoref{Eq:FullModel} we quote the form that the data model takes (for a derivation of the data model, see S16).
\begin{equation}
\label{Eq:FullModel}
\mathbfit{m} = \mathbfss{F}_{\mathrm{n}}^{-1} \mathbfss{P}\mathbfss{F}^{\prime}\left(\mathbfss{F}_z\mathbfit{a} + \mathbfss{Q}_z\mathbfit{q}\right) \ .
\end{equation}
Here, $\mathbfss{F}_z$ is a one-dimensional DFT matrix that models well-sampled fluctuations in the data on scales smaller than the bandwidth ($1/B \le \eta \le N_{c}/2B$, with $N_{c}$ the number of frequency channels and $B$ the bandwidth of the data from which the power spectrum is estimated). $\mathbfss{Q}_z$ is a quadratic model for large spectral scale fluctuations in the data, on scales longer than the bandwidth ($1/B \ge \eta$). The foregrounds are dominated by spectrally smooth emission. As a result, the majority of the power in the foregrounds will be absorbed by the quadratic model, preventing it from leaking into the well-sampled spectral scales of interest for estimating the power spectrum. However, we note that the primary purpose of the quadratic is simply to model large spectral scale fluctuations in the data to provide an unbiased estimate of power on the scales of interest, not specifically as a spectral model for the foregrounds. It was demonstrated in S16 that, for astrophysically realistic foreground simulations, while the majority of the power in the foregrounds is absorbed by the quadratic large spectral scale model, sufficient power remains to prevent unbiased recovery of the power spectrum of the EoR on large spatial scales. In \autoref{AMLSCM} we will describe the updated form that $\mathbfss{Q}_z$ takes in this work, which, in common with the quadratic model for large spectral scale structure, aims to minimise covariance with the EoR signal, but, in contrast, is explicitly informed by the intrinsic spectral structure of the foregrounds. $\mathbfss{F}^{\prime}$ encodes a two-dimensional Fourier transform of the model from the $uv$-domain to the image-domain. It consists of two model components at each frequency sampled by the data. In the first component, Fourier modes with spatial scales defined on a `coarse' grid with spacing $\sDelta\mathbfit{u}\simeq 1/\theta_{\mathrm{im}}$ model structure on spatial scales well sampled in the data. In the second component, a `sub-harmonic grid' with Fourier modes defined on a set of 10 log-uniformly spaced spatial scales between the size of the image and 10 times the size of the image is used to model large spatial scale structure not well sampled in the data, resulting, for example, in the case of Galactic foregrounds, from full-sky emission gradients towards the plane of the Galaxy. To speed-up the computation of the posterior, in this work we model diffuse foreground emission on angular scales up to the $13\fdg0$ field-of-view of our foreground simulations, enabling us to neglect the subharmonic grid in our analysis without introducing bias.\footnote{For the Nyquist-sampled $k$-space model for sub-40 m HERA baselines considered in this analysis, this reduces the number of $k$-space model parameters by a factor of $\sim2$ and the computation time of the posterior by a factor of $\sim8$.} $\mathbfss{P}$ is a matrix encoding the frequency-dependent primary beam response of the interferometer. $\mathbfss{F}_{\mathrm{n}}^{-1}$ is a non-uniform DFT matrix which transforms from the primary beam multiplied model sky to model visibilities at the frequency-dependent $uv$-coordinates sampled by the interferometer for the data set under consideration.

In addition to the data model describe above, we further assume that the redshifted 21 cm signal, from which the power spectrum is to be estimated, is spatially isotropic and, over the redshift interval under consideration, homogeneous (assuming the power spectrum of the 21 cm signal is approximately stationary) and uncorrelated between spatial scales\footnote{While the underlying hydrogen density distribution is expected to be well described as Gaussian at pre-reionization redshifts, it develops non-Gaussian features due to the formation of non-linear structures as reionization progresses. In this work we estimate the power spectrum of redshifted 21 cm emission from an EoR simulation for which the hydrogen neutral fraction is $x_{H_\mathrm{I}} \sim 0.88$. This corresponds to the relatively early stages of reionization where we expect an uncorrelated Gaussian model for the signal to be most reasonable. For an alternate approach using a non-Gaussian prior on the distribution of $k$-space amplitudes to jointly estimate higher order perturbations in the EoR signal with the power spectrum see Section 3.1 of S19.}. The covariance matrix $\sPsi$ of the $k$-space coefficients $\mathbfit{a}$ is given by,
\begin{equation}
\label{Eq:Prior}
\Psi_{ij} = \left< a(k_{i})a(k_{j})\right> = \varphi_{i}\delta_{ij},
\end{equation}
where $\delta_{ij}$ is the Kronecker delta function and $\varphi_{i}$ is the theoretical power spectrum of the EoR signal, in spherical annulus $i$, with units mK$^2$.

When constructing the spherically averaged power spectrum of the signal, we make use of the spatial homogeneity of the signal over a narrow redshift interval to average over spherical shells in $k$-space. We space the spherical shell limits logarithmically between the largest and smallest spatial scales sampled by the data set. The model for the spherically averaged power spectrum $\bm{\varphi}$ is given by a set of independent parameters $\varphi_{i}$, one for each $k$ bin $i$, where $k=\sqrt{k_x^2 + k_y^2 + k_z^2}$.

The final joint probability density of the model coefficients that define the power spectrum and the $k$-space signal coefficients is therefore,
\begin{equation}
\label{Eq:Prob}
\mathrm{Pr}(\bm{\varphi}, \mathbfit{a} , \mathbfit{q}\;|\; \mathbfit{d}) \; \propto \; \mathrm{Pr}(\mathbfit{d} | \mathbfit{q}, \mathbfit{a}) \; \mathrm{Pr}(\mathbfit{a} | \bm{\varphi}) \; \mathrm{Pr}(\bm{\varphi}) \; \mathrm{Pr}(\mathbfit{q}).
\end{equation}
As in S16, we assume a uniform prior on the amplitude of the large spectral scale parameters $\mathbfit{q}$, such that $\mathrm{Pr}(\mathbfit{q})=1$. In this work we consider the regime where the signal-to-noise ratio is sufficiently high to recover the power spectrum across the range of spectral scales accessible in the data (see \autoref{Noise} for details). In this regime, we select the least informative prior for our choice of $\mathrm{Pr}(\bm{\varphi})$: a log-uniform prior on the power spectral coefficients.

To implement these priors we sample from the parameter $\rho_{i}$, which parametrises $\varphi_{i}$, such that,
\begin{equation}
\label{Eq:PowerSpectrumPrior}
\varphi_{i} = \gamma(k_{i}) 10^{\rho_{i}}. 
\end{equation}
Here, $\gamma$ is a conversion factor between the dimensional power spectrum\footnote{Here, we refer to the dimensionless power spectrum as used in 21 cm astrophysics: $\Delta_{k}^{2} = \dfrac{k^{3}}{2\pi^{2}}P(k)$, where $P(k)$ has units $\mathrm{mK^{2}Mpc^{3}}$ and is defined by $\left\langle \widetilde{\sdelta T_\mathrm{b}}(\bm{k})\widetilde{\sdelta T_\mathrm{b}}^{*}(\bm{k}^\prime) \right\rangle \equiv (2\pi)^3\delta_D(\bm{k}-\bm{k}^\prime)P(\bm{k})$. In this equation, $\delta_D$ is the Dirac delta function, the angular brackets denote an ensemble average, $\widetilde{\sdelta T_\mathrm{b}}(\bm{k})$ is the Fourier transform of $\sdelta T_\mathrm{b}(\bm{x})$ and $\sdelta T_\mathrm{b}(\bm{x}) = (T_\mathrm{s}-T_\mathrm{r})(1-\exp(-\tau_\nu))/(1+z)$, with $\tau_\nu$, $T_\mathrm{s}$, $T_\mathrm{r}$, as the 21 cm optical depth, 21 cm spin temperature, and background radiation temperature at position $\bm{x}$, respectively. Correspondingly, the `dimensionless' 21 cm power spectrum considered here, rather than being dimensionless as is more commonly the case in cosmological contexts, has units of $\mathrm{mK^{2}}$.} $\bm{\varphi}$ and the dimensionless power spectral coefficients $10^{\bm{\rho}}$ (see S16 for details).

With this parametrisation, we can substitute $\mathrm{Pr}(\mathbfit{a}\vert\bm{\varphi})\mathrm{Pr}(\bm{\varphi}) = \mathrm{Pr}(\mathbfit{a}\vert\bm{\rho})\mathrm{Pr}(\bm{\rho})$ in \autoref{Eq:Prob}. From our log-uniform prior on $\bm{\varphi}$ we have $\mathrm{Pr}(\bm{\rho})=1$, which gives,
\begin{equation}
\label{Eq:LogUniformPrior}
\mathrm{Pr}(\mathbfit{a}\vert\bm{\rho})\mathrm{Pr}(\bm{\rho}) \propto \dfrac{1}{\sqrt{\mathrm{det}(\sPsi)}} \exp\left[ -\dfrac{1}{2}\mathbfit{a}^{\dagger}\sPsi^{-1}\mathbfit{a} \right] \ ,
\end{equation}
with $\sPsi$ as defined in \autoref{Eq:Prior}.

\subsection{Analytic marginalisation over the signal coefficients}
\label{MarginalisationOvertheSignalCoefficients}

Analytically marginalising over the signal coefficients $\mathbfit{a}$ and $\mathbfit{q}$ in \autoref{Eq:Prob} enables us to sample from the far smaller dimensional space of the power spectral coefficients $\bm{\varphi}$ and gives our final posterior probability distribution, $\mathrm{Pr}(\bm{\varphi}|\mathbfit{d})$. For a detailed derivation of this distribution from \autoref{Eq:Prob} see S19. Here, we quote the solution for the marginalised distribution for the case of log-uniform priors on the power spectral coefficients given in \autoref{Eq:LogUniformPrior},
\begin{eqnarray}
\label{Eq:Margin}
\mathrm{Pr}(\bm{\varphi}|\mathbfit{d}) &\propto& \frac{\mathrm{det} \left(\sSigma\right)^{-\frac{1}{2}}}{\sqrt{\mathrm{det} \left(\sPsi\right)~\mathrm{det}\left(\mathbfss{N}\right)}} \\
&\times&\exp\left[-\frac{1}{2}\left(\mathbfit{d}^\dagger\mathbfss{N}^{-1} \mathbfit{d} - \bar{\mathbfit{d}}^\dagger{\sSigma}^{-1}\bar{\mathbfit{d}}\right)\right]. \nonumber
\end{eqnarray}
Here, $\overline{\mathbfit{d}}=\mathbfss{T}^{\dagger}\mathbfss{N}^{-1}\mathbfit{d}$ is the projection of the weighted visibilities on the $k$-space grid of the model parameters, with $\mathbfss{T} = \mathbfss{F}_{\mathrm{n}}^{-1}\mathbfss{P}\mathbfss{F}^{\prime}(\mathbfss{F}_{z} + \mathbfss{Q}_{z})$ the matrix transform of the $k$-space parameters to visibilities described by \autoref{Eq:FullModel}, and $\sSigma= \mathbfss{T}^{\dagger}\mathbfss{N}^{-1}\mathbfss{T} + \sPsi^{-1}$ is the covariance matrix of $\overline{\mathbfit{d}}$.

\subsection{Astrophysically motivated large spectral scale model}
\label{AMLSCM}

In S19 a quadratic model is estimated jointly in the Fourier transform from frequency to the parameter $\eta$ in order to model any frequency variations that exist in the data that have periods longer than the bandwidth of the observation. It is demonstrated that, for simulated interferometric data derived from a simulated sky comprised of an EoR signal and a flat spectrum continuum foreground that is $10^{8}$ times greater in power, the inclusion in the data model of an accurate instrumental forward model, combined with a quadratic model for intrinsic spectral structure on spectral scales in excess of the 8 MHz bandwidth used in the analysis, is sufficient to recover unbiased estimates of the underlying EoR power spectrum on all well-sampled spatial scales in the data.

In S16 the same approach to power spectral estimation was used to recover the EoR power spectrum from data derived from simulated observations with HERA of a sky comprised of an EoR signal and a more realistic set of foreground simulations including: Galactic diffuse synchrotron emission (GDSE), synchrotron emission from extragalactic sources (EGS) incorporating a physically motivated model for synchroton self-absorption in the spectra in optically thick radio-loud AGN, and diffuse free-free emission resulting from the scattering of free electrons in diffuse \HII regions within the Galaxy.

With these more astrophysically realistic foreground simulations unbiased estimates of the EoR power spectrum on intermediate and small spatial scales ($\log_{10}(k[h\mathrm{Mpc^{-1}}]) > -0.80$) were recoverable. However, on larger spatial scales, foreground contamination dominated the recovered power spectral estimates. This demonstrates that, while including quadratics jointly estimated with Fourier basis vectors encoding the Fourier transform from frequency to the parameter $\eta$ in the analysis is useful as a generalised model for frequency variations in the data with periods longer than the bandwidth of the observation, recovery of unbiased power spectral estimates on the full range of spatial scales accessible to HERA requires an improved large spectral scale model for the foregrounds. In principle, cubic and higher order polynomial coefficients could be added to the quadratic large spectral scale model to absorb additional foreground power; however with increasing degrees of freedom, these will be increasingly covariant with the Fourier modes included in the EoR model. 

The total power in astrophysical emission at radio frequencies is dominated by GDSE and EGS synchrotron radiation. Synchrotron radiation is emitted through the acceleration of high energy cosmic-ray electrons in the magnetic field of a source. A synchrotron source with a power-law distribution of electron energies, $N(E) \propto E^{-p}$, where $N(E)$ is the number electrons at energy $E$, will exhibit a power-law spectrum of the form $\nu^{\beta}$, with temperature spectral index, $\beta = -(p-3)/2$ (e.g. \citealt{2011hea..book.....L}). Observationally, the spectral index distributions of the GDSE and EGS can be approximated by Gaussian distributions with means and standard deviations $<\beta>_\mathrm{GDSE} = -2.63$, $\sigma_{\beta_\mathrm{GDSE}} = 0.02$ and $<\beta>_\mathrm{EGS} = -2.82$, $\sigma_{\beta_\mathrm{EGS}} = 0.19$, respectively (see \autoref{SkyModels}). Free-free emission (thermal bremsstrahlung radiation) resulting from the scattering of free electrons in the warm ionised medium of the Galaxy is expected to account for approximately $1\%$ of the sky temperature at $150~\mathrm{MHz}$ \citep{1999A&A...345..380S}, but is nevertheless up to three orders of magnitude brighter than the EoR emission that we seek to detect. This diffuse gas is optically thin in the frequency range of interest for reionization experiments and has a well-determined power-law spectrum with temperature spectral index $\beta=-2.15$.

In the absence of an exact analytic model for foreground spectral structure along a given line of sight, the optimal large spectral scale model is one which can model the spectral structure of the foregrounds with sufficient accuracy to recover unbiased power spectral estimates, while being minimally covariant with the Fourier modes on the spatial scales of interest for recovering the power spectrum.

In this paper, we investigate the effectiveness of two astrophysically motivated parametrisation of the large spectral scale model:
\begin{enumerate}\item a constant plus single power law model (hereafter, the CPSPL model) of the form,
\begin{equation}
\label{Eq:LSSM1}
q_{0,j}+q_{1,j}(\nu/\nu_{0})^{b_{1}},
\end{equation}
with $q_{i,j}$ the amplitude of basis vector $i$ in $uv$-cell $j$. We consider two values of $b_{1}$: $b_{1} = <\beta>_\mathrm{GDSE}$ and $b_{1} = <\beta>_\mathrm{EGS}$, the mean spectral indices of the Galactic diffuse synchrotron emission and extragalactic source foreground emission, respectively, reflecting the spectra of the two most intense foreground components.
\item a constant plus double power law model (hereafter, the CPDPL model) of the form, 
\begin{equation}
\label{Eq:LSSM2}
q_{0,j}+q_{1,j}(\nu/\nu_{0})^{b_{1}}+q_{2,j}(\nu/\nu_{0})^{b_{2}},
\end{equation}
with $b_{1} = <\beta>_\mathrm{GDSE}$ and $b_{2} = <\beta>_\mathrm{EGS}$.
\end{enumerate}

When estimating the power spectrum of the EoR from an EoR plus foreground signal in \autoref{Analysis}, we will jointly estimate a model for the EoR and foregrounds in the manner described in \autoref{S16DataModel}. For all components of the power spectral analysis, we perform the analytic marginalisation described in \autoref{MarginalisationOvertheSignalCoefficients} and make use of \autoref{Eq:Margin} with $\mathbfss{T} = \mathbfss{F}_{\mathrm{n}}^{-1}\mathbfss{P}\mathbfss{F}^{\prime}(\mathbfss{F}_{z} + \mathbfss{Q}^{\prime}_{z})$, where, for each analysis, $\mathbfss{Q}^{\prime}_{z}$ encodes either one of astrophysically motivated large spectral scale models for the foregrounds, described above, or the quadractic large spectral scale model considered in previous work (see \autoref{Analysis} for details). In each case, we sample directly from the marginalised posterior for the spherical power spectrum coefficients $\mathrm{Pr}(\bm{\varphi}|\mathbfit{d})$ using nested sampling as implemented by the {\sc{MultiNest}} algorithm \citep{2009MNRAS.398.1601F}.

\section{Simulated data}
\label{SkyModels}

\subsection{The EoR signal}
\label{EoRSignal}

We use 21cmFAST (\citealt{2007ApJ...669..663M, 2011MNRAS.411..955M}) to generate simulations of the differential brightness temperature of the redshifted 21 cm signal as a function of redshift. We adopt the best-fitting cosmological parameter estimates from the Planck Collaboration XIII (\citealt{2016A&A...594A..13P}; Planck TT,TE,EE+lowP joint estimates). The timing of reionization, as modelled by 21cmFAST, is strongly influenced by three physical parameters: \begin{enumerate*} \item the UV ionising efficiency of high-z galaxies, $\zeta$, \item the mean free path of ionising photons within the ionised IGM, $R_\mathrm{mfp}$ (approximated in 21cmFAST as binary with a maximum horizon about ionizing sources at a radius $R=R_\mathrm{mfp}$), and \item the minimum virial temperature of halos hosting starforming galaxies, $T^\mathrm{min}_\mathrm{vir}$. \end{enumerate*} Estimates of these parameters are weakly constrained by current measurements. \citet{2017MNRAS.472.2651G} provide the following physically plausible ranges: $10 \lesssim \zeta \lesssim 250$, $5 \lesssim R_\mathrm{mfp} \lesssim 25~\mathrm{Mpc}$ and $10^{4} \lesssim T^\mathrm{min}_\mathrm{vir} \lesssim 10^{6}~\mathrm{K}$. Here, we select values: $\zeta=10.0$, $T^\mathrm{min}_\mathrm{vir}=10^{5}~\mathrm{K}$, $R_\mathrm{mfp}=22.2$, such that the neutral fraction is $x_{H_\mathrm{I}} \sim 0.88$ at redshift $z=7.7$, in agreement with the recent constraints on the history of reionization derived from Lyman-break Galaxies (\citealt{2019arXiv190109001H}). The corresponding observational central frequency of our cube is $\nu_{\mr{c}}=f_{21}/(1+ z)=163\ \mr{MHz}$, with $f_{21} \simeq 1420~\mathrm{MHz}$ the rest frame frequency of the 21 cm line.

We initialize 21cmFAST at $ z =300$ on a $2048^{3}$ grid with physical dimensions of one comoving $\mr{Mpc}^{3}$ per voxel and form the resulting brightness temperature cube on a $512^{3}$ lower resolution grid. These parameters correspond to a field of view with $\theta_{x}=\theta_{y}\approx 13\fdg0$ and a voxel angular and 21 cm spectral resolution of $\Delta \theta_{x}=\Delta \theta_{y}\approx 1.5~\mathrm{arcmin}$  and $\sDelta{\nu}\approx0.24\ \mr{kHz}$, respectively. The conversion from cosmological to observational units is given by (e.g. \citealt{2004ApJ...615....7M}),
\begin{eqnarray}
\label{CoordinateConversion2}
\sDelta \theta_{x}&=&\dfrac{\sDelta r_{x}}{D_\mathrm{M}( z )}\ , \nonumber \\
\sDelta \theta_{y}&=&\dfrac{\sDelta r_{y}}{D_\mathrm{M}( z )}\ , \\
\sDelta{\nu}&\approx&\dfrac{H_{0}f_{21}E( z )}{c(1+ z )^{2}} \sDelta r_{ z }\ , \nonumber 
\end{eqnarray} 
with, $D_{M}$ the transverse comoving distance from the observer to the redshift $z$ of the EoR observation (which for $\Omega_{\rm k}=0$, as assumed here, is simply equal to the comoving distance $D_{C}= (c/H_\mathrm{0})\int_{0}^{z} \mr{d}z^\prime / E(z^\prime)$, \citealt{1999astro.ph..5116H}) and $E(z) \equiv \sqrt{\Omega_{\rm M}(1+z)^{3} + \Omega_{\sLambda}}$ is the dimensionless Hubble parameter and $\sDelta r_{x}$, $\sDelta r_{y}$ and $\sDelta r_{z}$ are the transverse comoving separations of the cube at the redshift of the observation.

\begin{figure}
	\centerline{\includegraphics[width=\columnwidth]{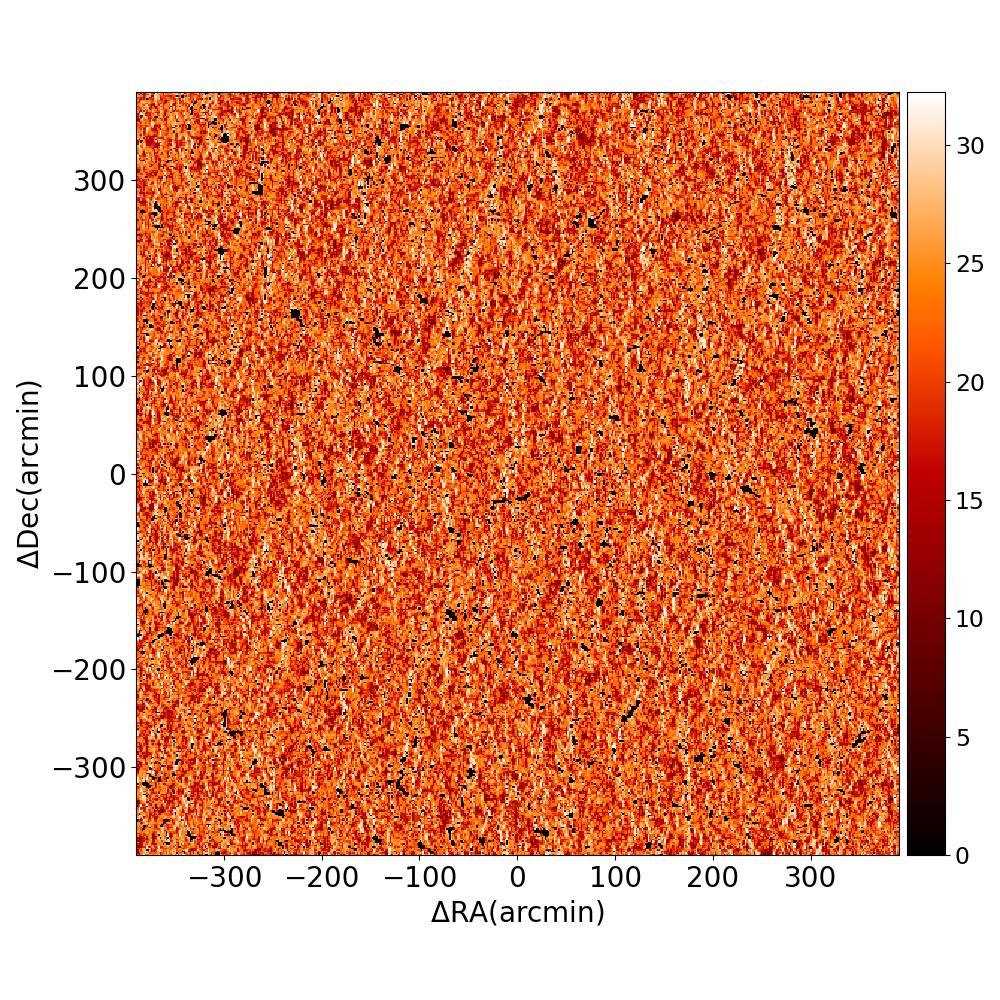}}
	\caption{The central channel, 163 MHz, of a 21cmFAST simulation of the differential brightness between the 21 cm spin temperature and CMB temperature at redshift $z=7.7$. The cube has a side of width 2048 $\mr{h^{-1}Mpc}$ and a neutral fraction, $x_{H_\mathrm{I}} \sim 0.88$. The colorscale is in mK.}
	\label{Signal}
\end{figure}

We estimate the power spectrum of the EoR under the assumption that it is stationary within the redshift range sampled by the data set. Cosmological isotropy means that the angular dependence of the signal analysed is not restricted by this assumption. However, the EoR signal undergoes temporal (and, correspondingly, spectral) evolution as the universe transitions from a neutral to an ionised state. The frequency bandwidth over which the evolution of the signal is expected to have a minimal impact on the power spectrum is of the order of $10~\mathrm{MHz}$ \citep{2012MNRAS.424.1877D, 2014MNRAS.442.1491D}. An analysis seeking to estimate the power spectrum of the EoR at a single point in its evolution is therefore restricted to estimation across a frequency interval within this bound. We select a 38 channel subset of our simulated EoR cube with a total bandwidth of $9.1~\mathrm{MHz}$. An image of the resulting EoR 21 cm emission simulation at 163 MHz is shown in \autoref{Signal}. 

\subsection{Galactic diffuse synchrotron emission}
\label{GDSE}

Emission from the Galaxy in the frequency range relevant to detection of the redshifted 21 cm signal from the EoR is dominated by Galactic diffuse synchrotron emission (GDSE). In S16 and S19 it was argued that large angular-scale structure resulting from, for example, the full sky emission gradient in GDSE towards the plane of the Galaxy has the potential to bias power spectral estimation, if not accounted for. In that analysis, a coarse plus sub-harmonic grid of $k$-space amplitude parameters was used to model power on Nyquist-criterion-fulfilling and large angular scales, respectively. The power spectrum was estimated from the coarse grid amplitudes on angular scales $\theta \lesssim 20\fdg0$ and power on angular scales in excess of this, which would otherwise leak into and bias the power spectra estimates, was modelled by the sub-harmonic grid parameters. 

Here, our focus is instead on the effectiveness of the large spectral scale model. We therefore construct our GDSE model such that power in the simulation is exclusively on angular scales smaller than the simulated field-of-view. In this case, power on scales modelled by the sub-harmonic grid large angular scale structure model is nullified, therefore consistent power spectral estimates will be recovered on well sampled scales irrespective of its inclusion. Thus, in this limit, we can elect not to include the sub-harmonic grid component of the data model without biasing our results.

To construct our GDSE simulations free from large spectral scale structure we adapt the approach used to construct a simulation of GDSE emission in \citet{2008MNRAS.389.1319J} (hereafter, J08). In the remainder of this section, we summarise the approach and include the details specific to our application; for a detailed description of that approach, see J08.

\begin{figure}
	\centerline{\includegraphics[width=\columnwidth]{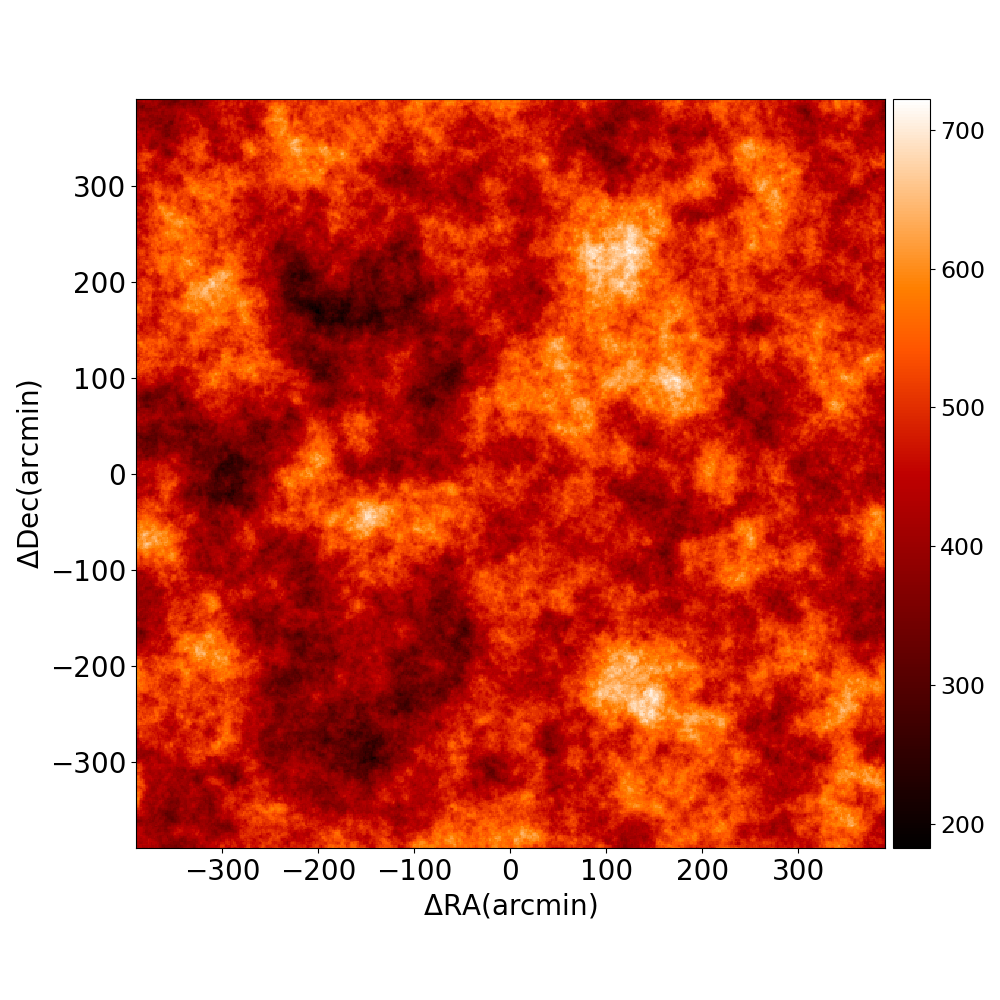}}
	\caption{The 163 MHz central channel of the high intensity GDSE region simulation, covering a $13\fdg0 \times 13\fdg0$ field of view. The emission is described by a power law two dimensional spatial power spectrum with an index of $\gamma=-2.7$ and is normalised to have a mean $\bar{T} = 471~\mathrm{K}$ and RMS $\sigma_\mathrm{T} = 63~\mathrm{K}$, at a resolution of $1\fdg0$ (see main text for details). The colorscale is in K.}
	\label{GDSEfig}
\end{figure}

We assume that the intensity and power law index of the GDSE can be spatially modelled by Gaussian random fields (GRFs). We construct our GRF realisation of the emission intensity field to have a power law two dimensional spatial power spectrum with a two dimensional power law index of $\gamma=-2.7$. We construct a four dimensional zero mean realisation of the Galactic diffuse synchrotron emissivity distribution, $J$, as,
\begin{equation}
\label{Eq:SynchEmissivity}
J(x,y,z,\nu) = J(x,y,z,\nu_{0})\left(\frac{\nu}{\nu_{0}}\right)^{\beta(x,y,z)},
\end{equation}
where we set reference frequency $\nu_{0} = 163~\mathrm{MHz}$ to the central frequency of our EoR signal simulation and $\beta(x,y,z)$ is a three dimensional realisation of the GDSE temperature spectral index distribution for which we set the mean and standard deviation to $<\beta>_\mathrm{GDSE} = -2.63$ and $\sigma_{\beta_\mathrm{GDSE}} = 0.02$, respectively, in agreement with values, given by \citet{2017MNRAS.464.4995M}, derived from measurements of the Galactic emission between 90 and 190 MHz with EDGES. We construct $J(x,y,z,\nu_{0})$ and $\beta(x,y,z)$ as $512^{3}$ voxel cubes with angular extent, $\theta_{x}=\theta_{y}\approx 13\fdg0$, and resolution, $\Delta \theta_{x}=\Delta \theta_{y}\approx 1.5~\mathrm{arcmin}$, such that $J(x,y,z,\nu)$, evaluated at 38 evenly spaced frequencies between 159 and 168 MHz, matches the extent and angular and spectral resolution of our EoR signal simulation. We obtain our three dimensional GDSE brightness temperature simulation, $T_\mathrm{b}(x,y,\nu)$, by integrating the emissivity cube along the line of sight,
\begin{equation}
\label{Eq:T_gdse}
  T_\mathrm{b}(x,y,\nu) = \bar{T} + C\int J(x,y,z,\nu)\mathrm{d}z \ .
\end{equation}
Here, $C$ is a normalization constant which we set with reference to the Global Sky Model (GSM; \citealt{2008MNRAS.388..247D}) evaluated at $\nu_{0}$. To normalise our GDSE simulation, we consider a $13\fdg0$ band centered on $\delta = -30\fdg0$, matching the $-30\fdg0$ latitude at which HERA is observing and spanning $360\fdg0$ in RA. We calculate the mean, $\bar{T}_{i}$, and variance, $\sigma_{T, i}^{2}$, of the brightness temperature distribution, at a pixel resolution of $\sim 1\fdg0$, in successive $13\fdg0$ fields, $i$, in right ascension. We exclude the most intense region of Galactic emission within $\pm 20\fdg0$ of the Galactic plane\footnote{The mean brightness temperature in this region is approximately an order of magnitude greater than the brightest region outside it, in the $13\fdg0$ band of sky centered on $\delta = -30\fdg0$ considered here.}. We approximate the sky area of interest for estimating the power spectrum of the EoR with HERA as being the remaining, non-excluded, region in the band of sky considered.
When constructing our GDSE simulations, we consider two cases: 
\begin{enumerate}
 \item In the first, which we henceforth will refer to as the high intensity GDSE region simulation, we set $C$ such that $T_\mathrm{b}(x,y,\nu)$ matches the RMS of the brightest region in this remaining band, with $\sigma_\mathrm{T} = 63~\mathrm{K}$, and set $\bar{T} = 471~\mathrm{K}$ to the mean brightness temperature in the corresponding field, to provide a conservative upper limit on the GDSE foregrounds that must be contended with when estimating the power spectrum of the EoR from HERA data. An image of the resulting GDSE emission simulation at 163 MHz is shown in \autoref{GDSEfig}.
 \item In the second, which we henceforth will refer to as the low intensity GDSE region simulation, we set $C$ such that $\sigma_\mathrm{T} = 30~\mathrm{K}$, which is approximately equal to the median RMS brightness temperature in the non-excluded region, and set $\bar{T} = 325~\mathrm{K}$ to the mean brightness temperature in the corresponding region.
\end{enumerate}

We note that a power law extrapolation of the $\sigma_\mathrm{T} = 1.3~\mathrm{K}$ RMS intensity of the GDSE emission, from $120~\mathrm{MHz}$, considered by J08, to the $163~\mathrm{MHz}$ central frequency of our simulations, yields $\sigma_\mathrm{T} \sim 0.45~\mathrm{K}$, assuming a temperature spectral index $\beta_\mathrm{GDSE} = -2.63$. The RMS normalisation of the low and high intensity GDSE regions described above are factors of approximately 70 and 150, respectively, higher than this normalisation and, indeed, the RMS of the coldest field in the latitude range considered here is also more than an order of magnitude higher than this value.

This difference can be understood as arising from the power law spatial power spectrum of GDSE, with power concentrated on large spatial scales, and the sensitivity as a function of spatial scale of the interferometers the simulations are designed for: LOFAR and HERA in J08 and this work, respectively. The concentration of sensitivity on shorter, sub-40 m, baselines in HERA and the absence of these short baselines in LOFAR means that HERA is sensitive to a significantly greater level of absolute GDSE power. See \autoref{ResolutionDependence} for a discussion of the correspondence between baseline lengths accessible to an instrument, and on which the power spectrum is estimated, and the expected level of power spectral contamination by foregrounds.

\subsection{Galactic diffuse free--free emission}
\label{FreeFree}

Thermal bremsstrahlung radiation resulting from the scattering of free electrons in diffuse \HII regions within the Galaxy is expected to account for approximately $1\%$ of the sky temperature at $150~\mathrm{MHz}$ \citep{1999A&A...345..380S}. \HII regions are optically thin in the frequency range of interest for reionization experiments and have a well determined power law spectrum with a temperature spectral index $\beta=-2.15$.

\begin{figure}
	\centerline{\includegraphics[width=\columnwidth]{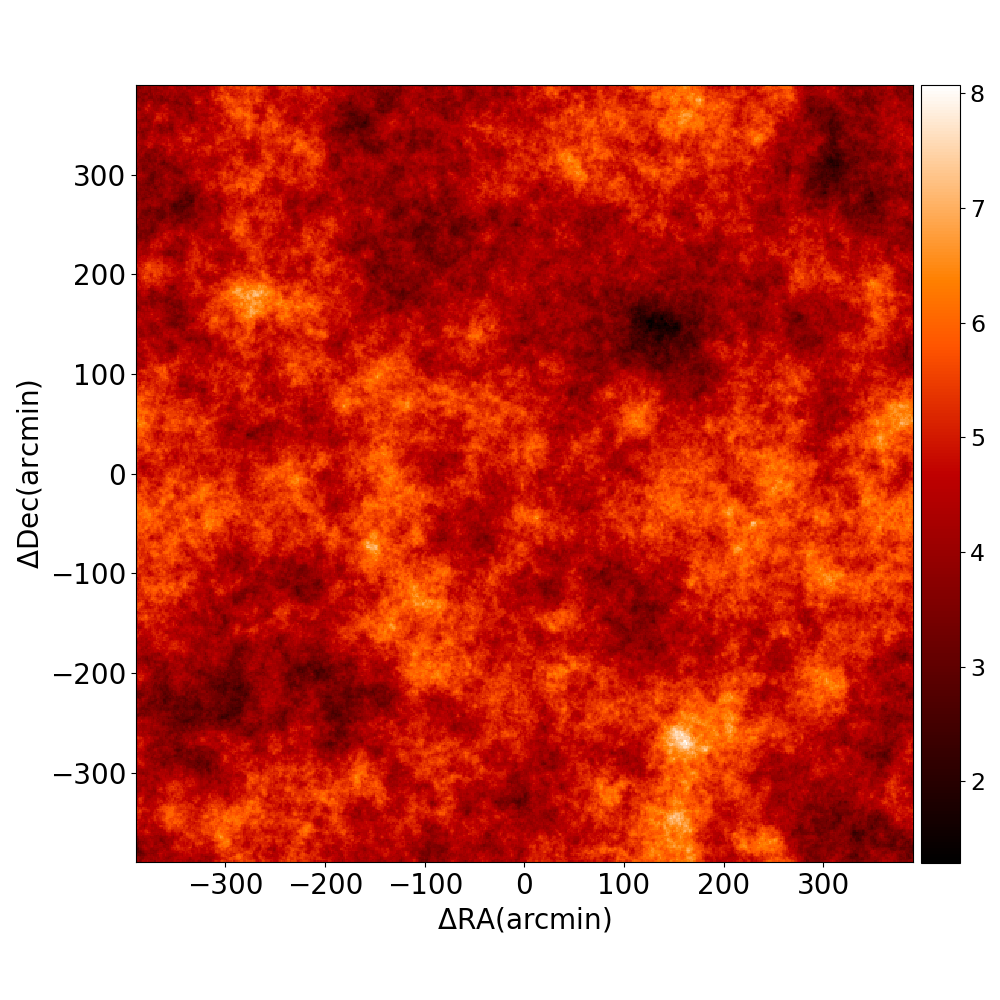}}
	\caption{The 163 MHz channel of the simulated diffuse free--free cube, covering a $13\fdg0 \times 13\fdg0$ field of view. The emission is described by a power law two dimensional spatial power spectrum with an index of $\gamma=-2.59$. The mean ($\bar{T}_\mathrm{T, ff} = 4.71~\mathrm{K}$) and variance ($\sigma_\mathrm{T, ff} = 0.63~\mathrm{K}$) of the emission is normalised to be one percent of their respective values in the GDSE simulation. The colorscale is in K.}
	\label{FreeFreefig}
\end{figure}

To construct our galactic diffuse free--free simulation, we apply the same procedure as described above for our GDSE simulation with the following modifications: \begin{enumerate*} \item we follow S16 and set the free--free spatial power spectral index as $\gamma = -2.59$, \item we fix the temperature spectral index to be $\beta_\mathrm{ff}=-2.15$ and \item we construct two free-free simulations, normalised such that their means and RMS brightness temperatures are $1\%$ of the corresponding values of the high and low intensity GDSE region simulations. \end{enumerate*} Our two resulting Galactic diffuse free--free emission simulations have $\sigma_\mathrm{T, ff} = 0.63~\mathrm{K}$, $\bar{T}_\mathrm{T, ff} = 4.71~\mathrm{K}$ and $\sigma_\mathrm{T, ff} = 0.30~\mathrm{K}$, $\bar{T}_\mathrm{T, ff} = 3.25~\mathrm{K}$, respectively. An image of the high intensity free--free emission simulation at 163 MHz is shown in \autoref{FreeFreefig}.

\subsection{Extragalactic source emission}
\label{EGSsim}

To construct our extragalactic source simulation we adopt the approach used to simulate extragalactic foregrounds in S16 and apply it to the $\sim10~\mathrm{MHz}$ band centred on 163 MHz relevant for the EoR signal simulation used in this work. For a detailed description of the approach, see \citet{2016MNRAS.462.3069S}. In brief, we use the Square Kilometre Array (SKA) Simulated Skies ($\mr{S^{3}}$) Simulation of Extragalactic Sources ($\mr{S^{3}}$-SEX; \citealt{2008MNRAS.388.1335W}) as the basis of our EGS simulation. For each source, we retrieve the radio flux at $151~\mr{MHz}$ and $610~\mr{MHz}$, the closest two frequencies to those of the desired $159$ -- $168~\mr{MHz}$ frequency range of our EGS simulation. We extrapolate the $\mr{S^{3}}$-SEX flux-densities from 151 MHz to 163 MHz on a per source basis using a power law of the form $S_{163}=S_{151} (163/151)^{-\alpha}$, with $\alpha = \log(S_{151}/S_{610})/\log(151/610)$. The full $\mr{S^{3}}$-SEX catalogue spans a $20\fdg0 \times 20\fdg0$ field. We take a $13\fdg0 \times 13\fdg0$ subset of the $163~\mr{MHz}$ catalogue to match the angular extent of our EoR cube.

We assign the spectral structure of the sources in our EGS simulation on a per-source basis. When doing this, we split the sources into two categories. Galaxies and AGN with rest frame synchrotron self-absorption turnover frequencies, $\nu_{\mr{t}} < 110(1+z)~\mr{MHz}$, are modelled as optically thin across the $159$ -- $168~\mr{MHz}$ frequency range of our foreground simulations. The remaining compact AGN, with an observed turnover frequency in excess of $110(1+z)~\mr{MHz}$, are modelled as optically thick\footnote{The rest-frame cut-off frequency derives from: $\nu_\mathrm{c} - W/2 = 110~\mr{MHz}$, with $\nu_\mathrm{c} = 163~\mr{MHz}$ the centre of the observed frequency band and taking $W \sim 100~\mathrm{MHz}$ as a typical synchrotron self-absorption turnover width (e.g. \citealt{2012MNRAS.424.2562Z}). For local ($z=0$) sources with rest-frame turnover frequencies $\nu_\mathrm{t}<\nu_\mathrm{tr}$, the impact of self-absorption on the observed spectrum will be small; it will be less still for sources at higher redshifts. See S16 for details.}. The spectra of those sources in the optically thin regime are calculated as power laws with spectral indices $\alpha$ drawn from the experimentally derived spectral index distribution of \citet{2014MNRAS.440..327L}, with a mean $\lan\alpha\ran=0.82$ and standard deviation $\sigma_{\alpha}=0.19$. For the remaining optically thick sources, we parametrise their spectra according to (e.g. \citealt{2011hea..book.....L}),
\begin{equation}
\label{Eq:SSASpectrum2}
S_\mr{ab}(\nu, p, l, B, \kappa) = \dfrac{S_{\nu}}{4\pi\chi_{\nu}}[1-\exp(-\chi_{\nu}l)].
\end{equation}
Here, $S_\mr{ab}$ is the synchrotron self-absorbed emission spectrum in the rest frame of the source, resulting from electrons in a randomly orientated magnetic field, $B$, with a power-law energy distribution of the form $N(E)\mr{d}E=\kappa E^{-p}\mr{d}E$, where $N(E)\mr{d}E$ is the number density of electrons in the energy interval $E$ to $E+\mr{d}E$ and $\kappa$ is the electron density distribution. $S_{\nu}$ is the optically thin power law spectrum, $l$ is the path length through the emitting region and $\chi_{\nu} = \chi_{0} \nu^{-(p+4)/2}$ is the synchrotron absorption coefficient with $\chi_{0} \propto \kappa B^{(p+2)/2}$. We generate spectra for sources falling in the optically thick regime according to \autoref{Eq:SSASpectrum2} in the manner described in S16.

We assume that the brightest extragalactic sources can be precisely characterised and, in the limit of negligible uncertainties on their amplitudes and positions, can be removed from the data in the visibilities. Additionally, in order to remain in the small $k$-cube regime, where sampling from the analytically marginalised power spectral posterior is computationally efficient, we follow the approach of S16 and restrict the baseline lengths on which we seek to estimate the power spectrum to those sampled with the greatest sensitivity by HERA. HERA is most sensitive on the shortest baselines and, as such, here, we bound our $k$-cube parameter space to sub-40 m baselines. When setting an upper bound on the highest flux-density sources included in our extragalactic simulations, we consider two cases:
\begin{enumerate}
 \item In the first, which we henceforth will refer to as the high intensity EGS simulation, we assume precise characterisation and removal of only those sources that will be resolved, at the highest frequency of our EGS simulation, in observations on the sub-40 m baseline lengths used in constructing the simulated data from which we seek to estimate the EoR power spectrum. That is, sources with flux densities above the (spatial resolution dependent) classical confusion noise limit corresponding to 40 m baselines.
We assume that a signal-to-noise ratio $q = S_0 / \sigma_{\rm c} = 5$ is required for the reliable detection of a point source with flux-density $S_0$, in a map with RMS confusion $\sigma_{\rm c}$. We estimate $\sigma_{\rm c}$ using a power law approximation to the differential source count distribution, $\mathrm{d}N/\mathrm{d}S = k S^{-\gamma}$, as (\citealt{1974ApJ...188..279C, 2012ApJ...758...23C}), 
\begin{equation}
\label{Eq:CondonConfusion}
\sigma = \left(\dfrac{q^{3-\gamma}}{3 - \gamma}\right)^{1/(\gamma-1)} (k \Omega_\mathrm{e})^{1/(\gamma - 1)} \ ,
\end{equation}
where  $\Omega_\mathrm{e} = \Omega_\mathrm{b}/(\gamma  -1)$ and $\Omega_\mathrm{b}$ is the solid angle of the synthesised beam. We take the \citet{2002ApJ...564..576D} power law fit of the high flux-density ($S>0.88~\mathrm{Jy}$) differential source count at 151 MHz, ($\mathrm{d}N/\mathrm{d}S(151~\mathrm{MHz}) = 4\times10^{3} (S/1~\mathrm{Jy})^{-\gamma}~\mathrm{Jy^{-1}sr^{-1}}$, with $\gamma=2.51$) and extrapolate to $163~\mathrm{MHz}$, assuming a mean source spectral index, $\alpha=0.82$. At $163~\mathrm{MHz}$, this gives, $\mathrm{d}N/\mathrm{d}S(163~\mathrm{MHz}) = 3.8\times10^{3} (S/1~\mathrm{Jy})^{-\gamma}~\mathrm{Jy^{-1}sr^{-1}}$. Using a maximum baseline length of $b_\mathrm{max} = 40~\mathrm{m}$ and a corresponding beam solid angle $\Omega_\mathrm{b} \simeq (\lambda/b_\mathrm{max})^{2}$, this results in an estimated confusion noise: $\sigma_\mathrm{c} = 8~\mathrm{Jy~beam^{-1}}$ and corresponds to a minimum flux-density for reliably detectable and individually countable point sources, $S>40~\mathrm{Jy}$. We consider this flux-density cut-off for sources included in our EGS simulation as a conservative upper limit on the flux-density of unsubtracted sources and source subtraction residuals in observations with HERA on sub-40 m baselines.
 \item In practice, observational data on longer baselines is already available from HERA and using these baselines will enable fainter sources below the confusion noise limit corresponding to 40 m baselines to be resolved and subtracted, even if we choose to restrict power spectral estimation to sub-40 m baselines. Additionally, the MWA GLEAM survey (\citealt{2015PASA...32...25W, 2017MNRAS.464.1146H}) has catalogued sources, in a $\sim 24\times10^{3}$ square degrees region, at declinations south of $30\fdg0$ and Galactic latitudes outside $10\fdg0$ of the Galactic plane at 20 frequencies between 72--231 MHz. The survey overlays the band of sky and frequency range observed by HERA and is $90\%$ cent complete at 170 mJy (\citealt{2017MNRAS.464.1146H}). As such, we consider a second case, which we henceforth will refer to as the low intensity EGS simulation, in which we assume that using a combination of longer HERA baselines, not contributing to our power spectral analysis, and sources catalogued in the GLEAM survey enables the removal of additional sources to a level such that the flux-density of unsubtracted sources and source subtraction residuals does not exceed $1~\mathrm{Jy}$. In this case, we include only sources with flux-densities $S \leq 1~\mathrm{Jy}$  in the simulation.
\end{enumerate}

\subsection{Instrument model}
\label{Noise}
 
To construct mock-observational data from the intrinsic sky simulations, described in the preceding subsections, we simulate their observation with a HERA-like instrument. For a detailed description of HERA see \citet{2017PASP..129d5001D}. 

We use a 331 antenna hexagonal close packed array configuration as an input to the Common Astronomy Software Applications ({\sc{CASA}}\footnote{http://casa.nrao.edu}) simobserve tool to obtain the set of sampled $(u,v)$ visibility coordinates corresponding to a 15 minute transit observation of our simulated sky and comprised of 30, 30 second integrations for 38, $\sim 240~\mathrm{kHz}$ channels spanning the frequency range 159--168 MHz. The computation time of the power spectrum posterior scales as number of model $k$-space voxels cubed; we therefore restrict our analysis to short baselines on which HERA is most sensitive and consider the baseline length subset: $2.5~\lambda < u < 22~\lambda$. We, additionally, coherently average over redundant baselines, increasing the average per-baseline sensitivity by a factor of $\sim18$.

We approximate the HERA primary beam $\mathbf{P}$ as a Gaussian with FWHM $9\fdg0 \nu_{i}/150.0~\mathrm{MHz}$ and we assume that the data under consideration has been perfectly calibrated.

We perform a non-uniform DFT of each channel of the zero-noise, primary beam multiplied model image to the frequency dependent $uv$-coordinates sampled during a 15 minute simulated observation, obtaining for each channel the sampled visibilities. We add uncorrelated white noise to the real and imaginary component of each of the sampled visibilities independently. The noise level on a visibility resulting from a pair of identical antennas individually experiencing equal system noise is given by (e.g. \citealt{1999ASPC..180.....T}),
\begin{equation}\label{VisabilityNoise}
\sigma_{V}=\dfrac{1}{\eta_{s}\eta_{a}}\dfrac{2k_{B}T_{\mathrm{sys}}}{A\sqrt{2\Delta\nu\tau}},
\end{equation}
\noindent
where $k_{B}=1.3806\times10^{-23}~\mathrm{JK^{-1}}$ is Boltzmann's constant, $T_{\mathrm{sys}}$ is the system noise temperature, $A$ is the antenna area, $\Delta\nu$ is the channel width, $\tau$ is the integration time and $\eta_{s}$ and $\eta_{a}$ are the system and antenna efficiencies, respectively.

Assuming system and antenna efficiencies of unity, an $150~\mathrm{m}^2$ area for a 14 m HERA antenna, a channel width of 240 kHz and a constant system noise temperature of $T_{\mathrm{sys}}=425~\mathrm{K}$, across our $9~\mathrm{MHz}$ bandwidth, yields a visibility noise $\sigma_{V}=2.1~\mathrm{Jy}$ per integration. We further assume a total observation length of 1000 hours, consisting of 4000 transits of the centre of the simulation cube, each of 15 minute duration\footnote{While, here, we use 4000 transits of the same field for simplicity, use of data from multiple fields will enable comparable results with fewer transits. We will address the application of our power spectral estimation framework to drift scan observations in future work.}. This yields an $0.033~\mathrm{Jy}$ effective noise on the data prior to redundant baseline averaging, which we add independently to the real and imaginary components of each of the sampled visibilities.

\section{Analysis}
\label{Analysis}

In this section we consider the dimensionless spherically averaged three-dimensional $k$-space intrinsic power spectrum of the EoR, recovered from simulated data comprised of EoR and foreground signals and noise (see \autoref{SkyModels}), using an updated version of the Bayesian analysis framework of S16 and S19, incorporating the astrophysically motivated large spectral scale models described in \autoref{AMLSCM}. For comparison, we additionally show power spectral estimates recovered when using a quadratic large spectral scale model, used in previous applications of the Bayesian analysis framework applied here.
 
When deriving the power spectral estimates presented here, we use a $k$-space model defined over the spatial resolution range $3\times10^{-3}~h\mathrm{Mpc^{-1}} < k_{\perp} < 2\times10^{-2}~h\mathrm{Mpc^{-1}}$, corresponding to the spatial resolution range spanned by the baseline lengths of our simulated data set. We simulate data sets comprised of EoR signal and one of four sets of foreground signals, which are defined as follows (see \autoref{SkyModels} for further details regarding the simulations):
\begin{itemize} \item the high-high simulation -- high intensity GDSE emission with RMS intensity at $163~\mathrm{MHz}$ and at $1\fdg0$ resolution of $\sigma_\mathrm{T} = 63~\mathrm{K}$ and bright EGS emission comprised of EGS up to a maximum flux-density of $S = 40~\mathrm{Jy}$, corresponding to the classical confusion noise limit of the sub-40 m baseline lengths on which the power spectrum is to be estimated;
\item the high-low simulation -- high intensity GDSE emission and low intensity EGS emission that assumes that sources resolved on the maximum baseline lengths present in HERA 331 and bright sources catalogued by the GLEAM survey (see \autoref{EGSsim}) have been subtracted from the data set to a level such that remaining sources and source subtraction residuals do not exceed flux densities of $S = 1~\mathrm{Jy}$;  
\item the low-high simulation -- low intensity GDSE and high intensity EGS emission and  
\item the low-low simulation -- low intensity GDSE and low intensity EGS emission. 
\end{itemize}
In each of the above cases, we additionally include free-free emission with an RMS intensity equal to 1\% of the RMS intensity of the simulated GDSE emission in the case under consideration.

\begin{figure*}
	\centerline{
	\includegraphics[width=0.50\textwidth]{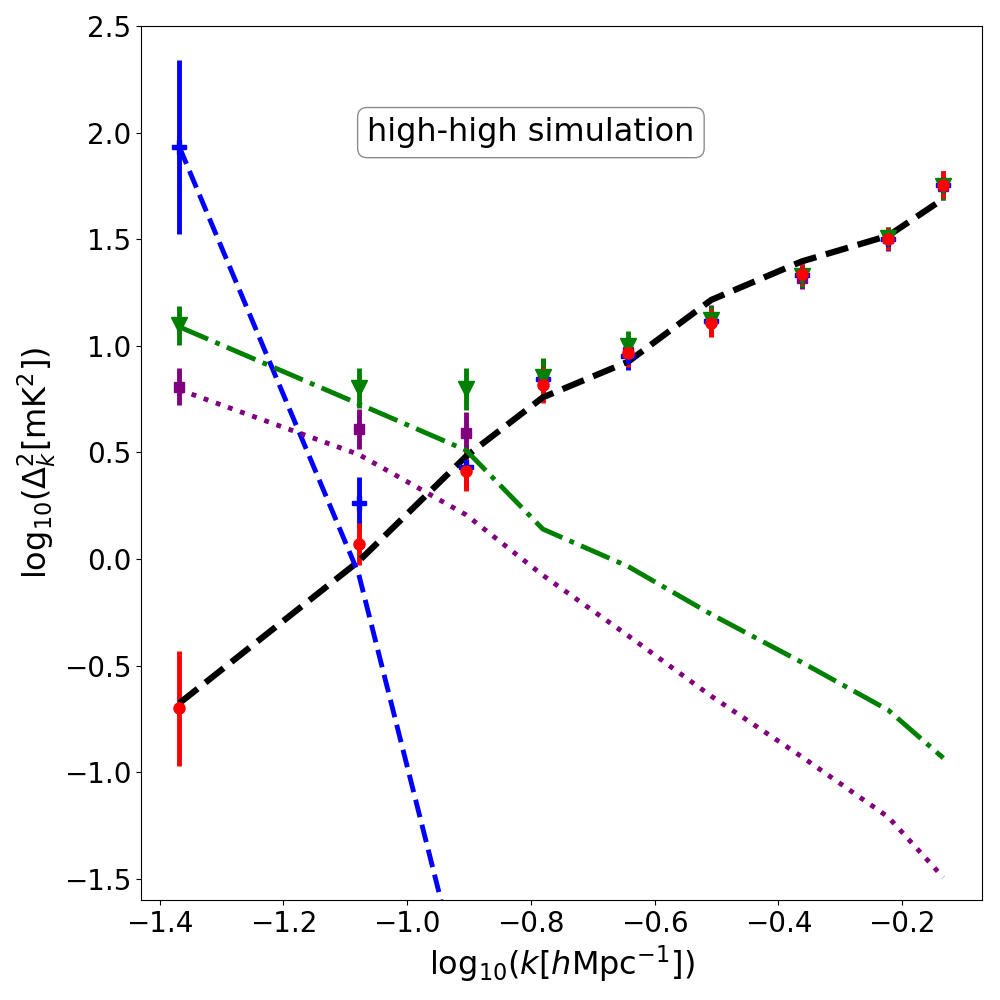}
	\includegraphics[width=0.50\textwidth]{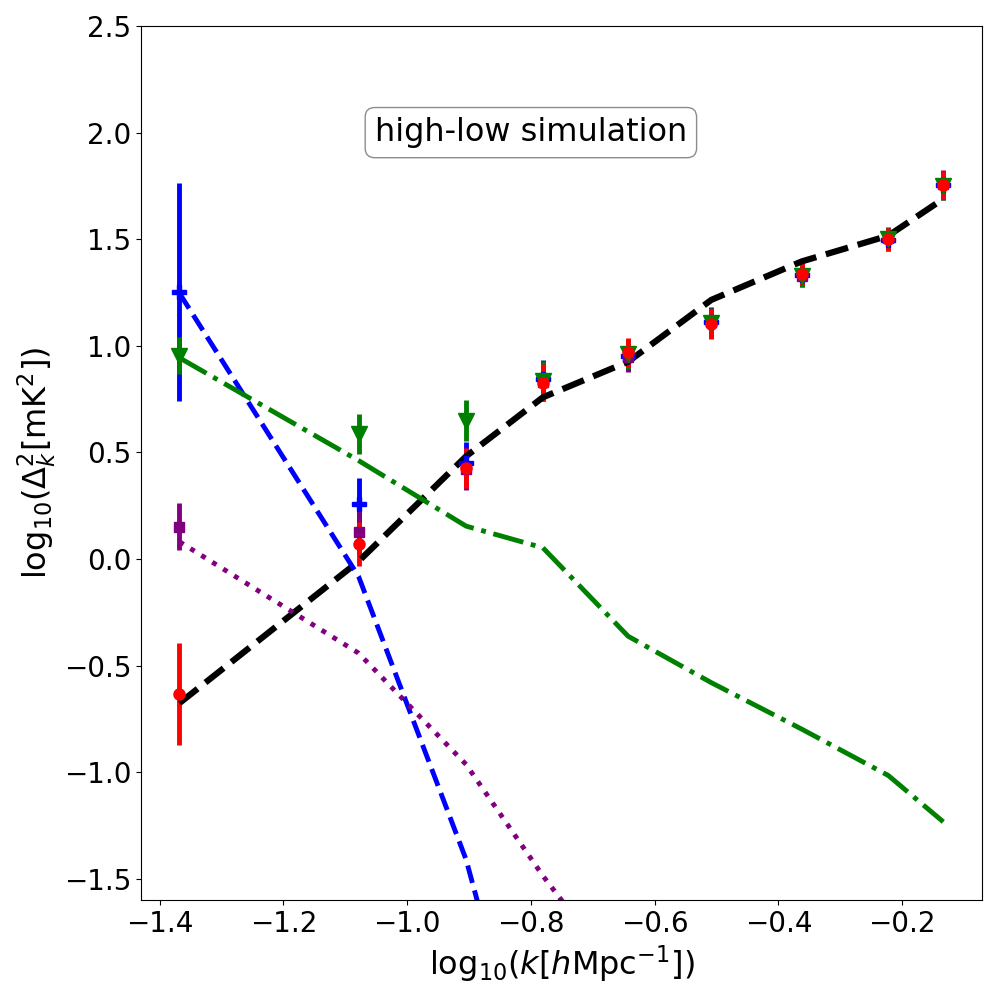}}
	\centerline{
	\includegraphics[width=0.50\textwidth]{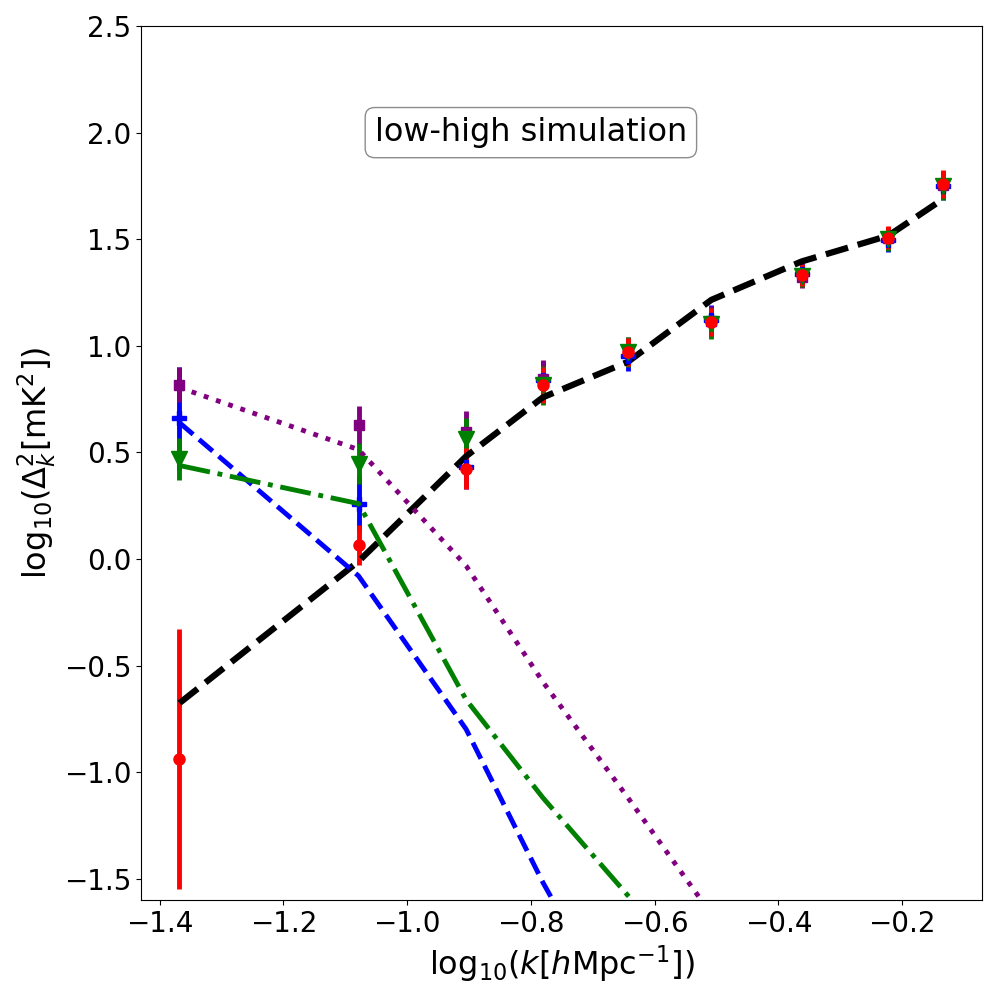}
	\includegraphics[width=0.50\textwidth]{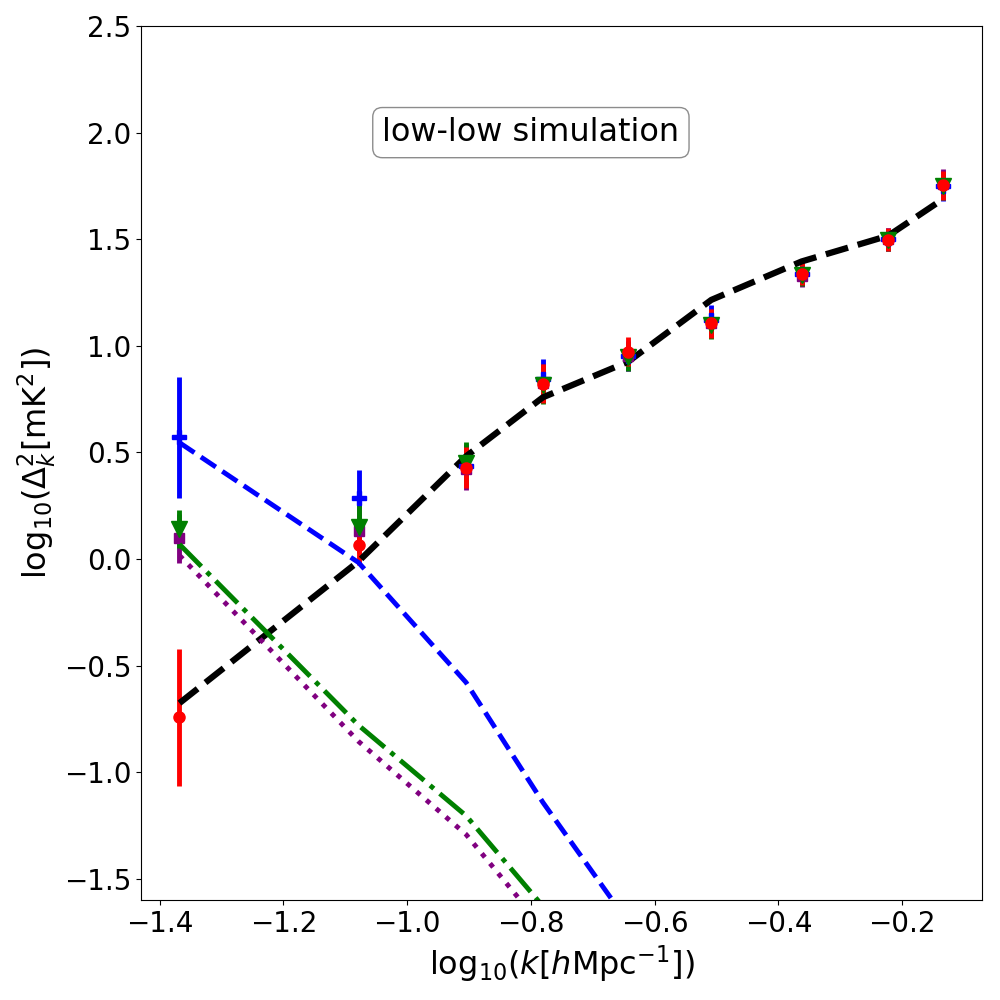}}
	\centerline{
	\includegraphics[width=1.0\textwidth]{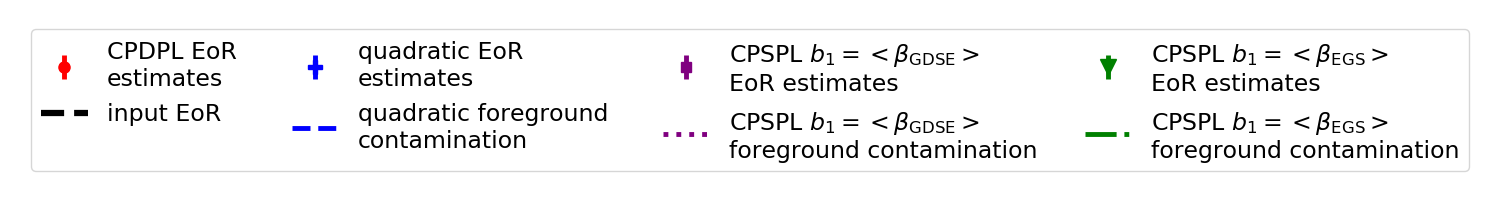}}
	\caption{Input (black thick dashed line) and recovered values with four large spectral scale models (points with one sigma error bars) of the spherically averaged dimensionless power spectrum of the EoR. Recovered values are estimated from simulated data comprised of an EoR signal and four sets of foreground simulations, with the foreground simulation used in the analysis displayed in the top right of each figure. In each case, the mean posterior EoR power spectrum estimates and one sigma uncertainties recovered with four models for the large spectral scale structure in the data are shown. The marker types and colours used for the different models are consistent between plots and are displayed in the legend.The power spectral contamination by foregrounds with the quadratic and CPSPL large spectral scale model with power law index $b_{1} = <\beta>_\mathrm{GDSE}$ and with $b_{1} = <\beta>_\mathrm{EGS}$ are shown are shown in the thin dashed, dotted and dot-dashed lines, respectively with colours matched to the corresponding mean posterior EoR power spectral estimates. There is no detectable power spectral contamination by foregrounds when using the CPDPL model and, as such, the corresponding line is not present.}
	\label{SphPS_LWSM_comparison}
\end{figure*}

\subsection{Spherically averaged power spectrum}
\label{SPS}
In \autoref{SphPS_LWSM_comparison}, we show the input (black thick dashed line) and recovered values (points with one sigma error bars) of the spherically averaged dimensionless power spectrum of the EoR, estimated from simulated data comprised of an EoR signal and, clockwise from top left, the high-high, high-low, low-high and low-low foreground simulations, respectively. In each case, we display the recovered power spectral estimates derived with a data model in which a Fourier mode model for structure in the EoR signal on Nyquist criterion fulfilling spatial scales in the data set is jointly estimated with one of the four large spectral scale models described in \autoref{AMLSCM}.
It can be seen that in the limit of no uncertainty on the forward model of the instrument and across the $9~\mathrm{MHz}$ band in which we conduct our analysis, all of the large spectral scale models considered are sufficiently good descriptions of the intrinsic spectral structure of the foregrounds to enable recovery of the EoR power spectrum on intermediate and small spatial scales ($\log_{10}(k[h\mathrm{Mpc^{-1}}]) > -0.80$) in all .

The results recovered using a quadratic large spectral scale model are qualitatively in agreement with those of S16 where it was found that, in both a bright GDSE region overlaying the plane of the Galaxy and in cold regions of the Galaxy lying out of the plane of the Galaxy, where GDSE (the most significant foreground contaminant on short baselines, where HERA in most sensitive) is least intense, a quadratic large spectral scale model is not sufficient to recover unbiased power spectral estimates on large spectral scales.

It can be seen that the CPSPL model, for both choices of power law index, perform comparably to, or better than, the quadratic large spectral scale model for the foregrounds with respect to mitigating bias in the recovered EoR power spectrum estimates at $\log_{10}(k[h\mathrm{Mpc^{-1}}]) \simeq -1.4$ but, nevertheless remain contaminated in each of the foreground scenarios considered. Marginally consistent estimates (correct to within two sigma) of the EoR power spectrum at $\log_{10}(k[h\mathrm{Mpc^{-1}}]) \sim -1.0$ are recovered when using the CPSPL large spectral scale models, with power law index given by the mean spectral index of the GDSE emission when analyzing the data including the high-low foreground simulation. Similarly consistent estimates are recovered when the large spectral scale model power law index is given by either of the mean spectral indices of the GDSE or EGS components of the foregrounds when analyzing the data set using low-low foreground simulation.

In contrast, when analyzing either of the data sets including high EGS foregrounds (high-high and low-high simulations), estimates of the EoR power spectrum are significantly contaminated at $\log_{10}(k[h\mathrm{Mpc^{-1}}]) \sim -1.0$ when utilising the CPSPL large spectral scale model with either choice of model power law index. This derives from two effects. Firstly, the ratio of the total power in the high and low GDSE foregrounds is approximately 5, where as for the EGS simulations it is approximately 9.
As a result, between the high and low EGS models, foreground contamination that results from EGS is reduced  by a factor that exceed the equivalent reduction in foreground contamination in GDSE between the high and low GDSE models. Secondly, the width of the spectral index distribution that describes the EGS population is wider than that which describes GDSE. As such, the CPSPL large spectral scale model with power law index $b_{1} = <\beta>_\mathrm{GDSE}$ provides a better description of the GDSE component of the foregrounds than the equivalent model with $b_{1} = <\beta>_\mathrm{EGS}$ describes the EGS population. 
Further evidence for this can be seen in the comparable levels of contamination at low-$k$, when utilising the CPSPL large spectral scale model with $b_{1} = <\beta>_\mathrm{GDSE}$, when including the high-high or low-high foreground simulations, where GDSE and free-free emission are reduced, but EGS emission is held fixed, which demonstrates that the GDSE and free-free emission components of the foregrounds are well modelled and that the contamination present is dominated by EGS. In contrast, the level of contamination of the recovered power spectral estimates when using the CPSPL model with $b_{1} = <\beta>_\mathrm{EGS}$ depends to a greater extent on the intensity of both the GDSE and EGS foregrounds, such that neither is sufficiently well modelled to not contribute to foreground contamination of the large spatial scale modes of the EoR power spectrum with this model. Nevertheless, a reduction in the level of GDSE emission between high-high and low-high has a greater impact on the total level of foreground contamination with this model than the corresponding reduction in EGS intensity between simulations high-high and high-low foreground simulations, demonstrating that the contribution of GDSE emission to the total power spectral contamination when using the CPSPL model with $b_{1} = <\beta>_\mathrm{EGS}$ exceeds that of EGS, as may be expected.

When using any one of the three models discussed above, recovered estimates of the power spectrum on the two largest $k$-scales ($\log_{10}(k[h\mathrm{Mpc^{-1}}]) \simeq -1.4$ and $-1.0$) are contaminated by foreground to varying degrees, with, at best, marginally consistent detection of the $\log_{10}(k[h\mathrm{Mpc^{-1}}]) \simeq -1.0$ mode in data sets with extragalactic sources and source subtraction residuals not exceeding $S = 1~\mathrm{Jy}$ at $163~\mathrm{MHz}$, but with foregrounds contributing at least 80\% of the recovered power at $\log_{10}(k[h\mathrm{Mpc^{-1}}]) \simeq -1.4$ in all cases.
In contrast, the CPDPL large spectral scale model outperforms each of these by a significant margin, with respect to recovery of unbiased estimates of the power spectrum on large spatial scales. When jointly estimating a model for the EoR signal CPDPL model, we find that the power spectral contamination by foregrounds is undetectable at the signal to noise level considered here, for all of the levels of foregrounds contamination analysed. This enables recovery of power spectral estimates that are consistent with the underlying EoR power spectrum on all of the spatial scales accessible in the data set in each of the foreground scenarios considered.
This improved recovery is significant for both power spectral bins at $\log_{10}(k[h\mathrm{Mpc^{-1}}]) < -0.8$, but is particularly evident in the recovered estimates centered on $\log_{10}(k[h\mathrm{Mpc^{-1}}]) \sim -1.4$, in which the power is greater than 99\% and 90\% comprised of foreground contamination and has greater than five sigma deviations from the underlying EoR power spectrum when using the quadratic and CPSPL large spectral scale models, respectively, when estimated in the presence of intense GDSE, EGS and free-free emission included in the high-high foreground simulation. Even in the more optimistic low-low foreground scenario, foreground contamination comprises greater than 90\% and 80\% of the recovered power and there are greater than 3 sigma deviations from the underlying EoR power spectrum when using the quadratic and CPSPL large spectral scale models, respectively. In contrast, power spectral estimates consistent with the underlying EoR power spectrum, to their one sigma uncertainties, are recovered with the CPDPL model in both cases.

\subsection{The resolution dependence of foreground contamination}
\label{ResolutionDependence}

When constructing our simulated GDSE foregrounds in \autoref{GDSE}, we use the approach described in J08 for simulating Galactic diffuse synchrotron emission. As a result of the different angular scales probed by HERA and LOFAR, for which the simulations here and in J08 are constructed, respectively, we derive RMS normalisations for our low and high intensity GDSE region simulations from the GSM at $1\fdg0$ resolution, that are larger than that used in J08 by factors of 70 and 150, respectively (following power-law extrapolation of the RMS of their simulations at $120~\mathrm{MHz}$ to the $163~\mathrm{MHz}$ central frequency considered here; see \autoref{GDSE} for details).

The primary cause of this difference in normalisation is the difference in the brightness of the Galaxy on the differing baseline lengths on which HERA and LOFAR are most sensitive to the EoR signal. HERA is most sensitive to the signal on its shortest $14.6~\mathrm{m}$ baselines. In contrast, LOFAR is most sensitive at baseline lengths of $\mathcal{O}(100)~\mathrm{m}$. The power law spatial power spectrum of the GDSE (see e.g. \citealt{2008A&A...479..641L}) means that power in the GDSE emission is concentrated on large angular scales sampled by shorter baselines. As such, LOFAR's concentration of sensitivity on smaller angular scales\footnote{If $P(k)$, the three dimensional $k$-space power spectrum of the EoR or power spectral contamination by the foregrounds, is assumed to be spatially separable, we can write, $P(k_{\perp}, k_{\parallel}) = P_{k_{\perp}}(k_{\perp})P_{k_{\parallel}}(k_{\parallel})$ and it follows that $P_{k_{\perp}} \propto P_{uv}$, with $P_{uv}$ the two dimensional spatial power spectrum. In this case, the optimal baseline lengths for estimating the EoR signal can be determined for a given EoR signal and foreground model by calculating the ratio of the two dimensional spatial power spectrum of the EoR and foregrounds as a function of spatial scale. In S16 an approximate wavelength range $35~\lambda \le \abs{\mathbfit{u}} \le 55~\lambda$ at $126~\mathrm{MHz}$ is found to be the optimal spatial scale range, for the foreground models considered there, for a wide variety of EoR models (see S16 for details). At $120~\mathrm{MHz}$, this corresponds to baselines in the approximate range $80 \le \abs{b} \le 130~\mathrm{m}$. Outside of this range, a greater demand is placed on the effectiveness of techniques for separating the EoR and foreground signals.} has the benefit that the GDSE foreground observed by LOFAR is far less intense than that observed by HERA and, correspondingly, significantly decreases the level of contamination due to GDSE that will result from a given foreground spectral structure.

Furthermore, in the EGS simulations of J08, extragalactic point sources exceeding a flux-density of $10~\mathrm{mJy}$ are excluded. In contrast, in the high and low intensity EGS simulations, relevant to HERA, considered here, sources up to a maximum flux density of $S \le 40~\mathrm{Jy}$ and $S \le 1~\mathrm{Jy}$ are included, respectively (see \autoref{EGSsim} for details). This difference reflects the greater number of sources that are well resolved and potentially removable from the data to high precision owing to the significantly improved imaging resolution with LOFAR relative to HERA. 

We note that with the far lower intensity of GDSE and EGS emission used in the simulations of J08, the power spectral contamination in estimates of the EoR power spectrum with the quadratic and CPSPL large spectral scale models would be greatly reduced. Assuming a reduction in power spectral contamination proportional to the square of the reduction in the RMS intensity of the foregrounds, in this regime,  both the quadratic and CPSPL parametrisations for the large spectral scale model can be predicted to be sufficient to model the intrinsic spectral structure in the foregrounds to a sufficient level for there to be negligible bias in corresponding estimates of the EoR power spectrum.

In this case, while, of the foreground models considered here, the CPDPL large spectral scale model outperforms the alternatives, the additional degree of freedom relative to the CPSPL model results in an increased degree of correlation between the large spectral scale model and the long wavelength Fourier mode components of the data model, and correspondingly larger uncertainties on the power spectrum at low-$k$. As such, for sufficiently low foregrounds, this reduction in power spectral uncertainties means a CPSPL model can be competitive with or improve on the estimates of the EoR power spectrum recoverable with the CPDPL model.

\section{Summary}
\label{Conclusions}

Building on the Bayesian power spectral estimation methodology introduced in S16 and S19, we have shown that, by adapting the large spectral scale model to better incorporate a priori knowledge of the spectral structure of the most intense foreground components, we can significantly reduce bias in recovered estimates of the EoR power spectrum relative to using a generic polynomial model as applied in S16 and S19. We have investigated the use of a constant plus single power law (CPSPL) large spectral scale model of the form: $q_{0,j}+q_{1,j}(\nu/\nu_{0})^{b_{1}}$, both for $b_{1} = <\beta>_\mathrm{GDSE}$ and $b_{1} = <\beta>_\mathrm{EGS}$ and a constant plus double power law (CPDPL) large spectral scale model of the form: $q_{0,j}+q_{1,j}(\nu/\nu_{0})^{b_{1}}+q_{2, j}(\nu/\nu_{0})^{b_{2}}$ with $b_{1} = <\beta>_\mathrm{GDSE}$ and $b_{2} = <\beta>_\mathrm{EGS}$. In both models, $q_{i,j}$ is the amplitude of basis vector $i$ in $uv$-cell $j$ and $<\beta>_\mathrm{GDSE} = -2.63$ is the mean temperature spectral index of the GDSE component of the foregrounds and $<\beta>_\mathrm{EGS} = -2.82$ is the mean temperature spectral index of the EGS component of the foregrounds.

We have constructed foreground simulations comprised of Galactic diffuse synchrotron emission, extragalactic sources and diffuse free-free emission from the Galaxy in the 159--168 MHz frequency range and analysed data sets comprised of simulated observations, of 1000 h duration, of these foregrounds and a simulated EoR signal at redshift $z=7.7$, on sub-40 m baselines of HERA in 331 antenna configuration. When constructing the data sets we have considered four levels of foreground contamination:
\begin{itemize*} \item high intensity GDSE emission with RMS intensity of $\sigma_\mathrm{T} = 63~\mathrm{K}$, at $163~\mathrm{MHz}$ and at $1\fdg0$ resolution, and bright EGS emission comprised of EGS up to a maximum flux-density of $S = 40~\mathrm{Jy}$,
\item high intensity GDSE emission and low intensity EGS emission that assumes that sources and source subtraction residuals do not exceed flux densities of $S = 1~\mathrm{Jy}$,  
\item low intensity GDSE and high intensity EGS emission and  
\item low intensity GDSE and low intensity EGS emission. 
\end{itemize*}

We find that in the limit of no uncertainty on the forward model of the instrument and across the $9~\mathrm{MHz}$ band in which we conduct our analysis, all of the large spectral scale models considered are sufficiently good descriptions of the intrinsic spectral structure of the foregrounds to enable recovery of the EoR power spectrum on intermediate and small spatial scales ($\log_{10}(k[h\mathrm{Mpc^{-1}}]) > -0.80$) in all four foreground scenarios. The CPSPL model, for both choices of power law index, perform comparably to, or better than, the quadratic large spectral scale model for the foregrounds with respect to mitigating bias in the recovered EoR power spectrum estimates at $\log_{10}(k[h\mathrm{Mpc^{-1}}]) \simeq -1.4$, but, nevertheless, remain contaminated.

In contrast, when jointly estimating a model for the EoR signal with the astrophysically motivated CPDPL parametrisation of the large spectral scale structure model, we recover power spectral estimates that are consistent with the underlying EoR power spectrum on all of the spatial scales accessible in the data set for all of the foreground scenarios considered. As such, use of the astrophysically motivated CPDPL parametrisation for the large spectral scale power for the foregrounds significantly improves performance with respect to previous applications of our Bayesian power spectral estimation framework and expands the $k$-space volume accessible for recovery of unbiased estimates of the EoR power spectrum and derived astrophysical parameter constraints. Furthermore, this is achieved without increasing the model complexity, and corresponding uncertainty on the power spectral estimates, relative to the quadratic large spectral scale used in earlier work.

\subsection{Future work}
\label{Future work}
The level of foreground contamination and its statistical significance is a function of the intensity of the foregrounds in the data. This, itself, is a function of the field observed and frequency range of the data set. For the foreground intensities and signal-to-noise level considered in this analysis, we find that our updated large spectral scale model with power law indices matched to the mean spectral indices of the GDSE and of the EGS emission is sufficient for no statistically significant foreground contamination to be detectable. However, as 21 cm experiments push towards CD, at lower frequencies, the foreground component of the sky signal increases by up to an order of magnitude in intensity. In the presence of more intense foregrounds, testing the optimality of our choice of power law indices in our large spectral scale model or, if necessary, determining the optimal parameter values, will be of increased importance. Additionally, in a more realistic scenario, the EoR power spectrum will be estimated from a data set where there is imperfect knowledge of the spectral structure of the foregrounds. In this case, rather than assigning a value to the power law index, a preferred approach would be to estimate the optimal power law index from the data. A Bayesian evidence based analysis can provide a rigorous foundation for testing the optimality of our choice of power law indices and, in the case that it is not, for deriving the optimal power law indices describing the foregrounds present in a given data set. We will consider an extension of this type, to the analysis carried out in this paper, in upcoming work.

\section*{Acknowledgements}

PHS and JCP both acknowledge support from NSF award \#1636646  and Brown University's Richard B. Salomon Faculty Research Award Fund. This research was conducted using computational resources and services at the Center for Computation and Visualization, Brown University. PHS thanks Irina Stefan for valuable discussions and helpful comments on a draft of this manuscript. 



\label{lastpage}


\begin{thebibliography}{99}

\bibitem[\protect\citeauthoryear{Barkana \& Loeb}{2007}]{2007RPPh...70..627B} Barkana R., Loeb A., 2007, RPPh, 70, 627 

\bibitem[\protect\citeauthoryear{Bonaldi \& Brown}{2015}]{2015MNRAS.447.1973B} Bonaldi A., Brown M.~L., 2015, MNRAS, 447, 1973 

\bibitem[\protect\citeauthoryear{Bowman et al.}{2018}]{2018Natur.555...67B} Bowman J.~D., Rogers A.~E.~E., Monsalve R.~A., Mozdzen T.~J., Mahesh N., 2018, Natur, 555, 67 

\bibitem[\protect\citeauthoryear{Bowman, Morales \& Hewitt}{2009}]{2009ApJ...695..183B} Bowman J.~D., Morales M.~F., Hewitt J.~N., 2009, ApJ, 695, 183 

\bibitem[\protect\citeauthoryear{Chapman et al.}{2012}]{2012MNRAS.423.2518C} Chapman E. et al., 2012, MNRAS, 423, 2518 

\bibitem[\protect\citeauthoryear{Chapman et al.}{2013}]{2013MNRAS.429..165C} Chapman E., et al., 2013, MNRAS, 429, 165 

\bibitem[\protect\citeauthoryear{Condon}{1974}]{1974ApJ...188..279C} Condon J.~J., 1974, ApJ, 188, 279 

\bibitem[\protect\citeauthoryear{Condon et al.}{2012}]{2012ApJ...758...23C} Condon J.~J., et al., 2012, ApJ, 758, 23 

\bibitem[\protect\citeauthoryear{Datta et al.}{2014}]{2014MNRAS.442.1491D} Datta K.~K., Jensen H., Majumdar S., Mellema G., Iliev I.~T., Mao Y., Shapiro P.~R., Ahn K., 2014, MNRAS, 442, 1491 

\bibitem[\protect\citeauthoryear{Datta et al.}{2012}]{2012MNRAS.424.1877D} Datta K.~K., Mellema G., Mao Y., Iliev I.~T., Shapiro P.~R., Ahn K., 2012, MNRAS, 424, 1877 

\bibitem[\protect\citeauthoryear{Datta et al.}{2012}]{2012MNRAS.424..762D} Datta K.~K., Friedrich M.~M., Mellema G., Iliev I.~T., Shapiro P.~R., 2012, MNRAS, 424, 762 

\bibitem[\protect\citeauthoryear{de Oliveira-Costa et al.}{2008}]{2008MNRAS.388..247D} de Oliveira-Costa A., Tegmark M., Gaensler B.~M., Jonas J., Landecker T.~L., Reich P., 2008, MNRAS, 388, 247 

\bibitem[\protect\citeauthoryear{DeBoer et al.}{2017}]{2017PASP..129d5001D} DeBoer D.~R., et al., 2017, PASP, 129, 045001 

\bibitem[\protect\citeauthoryear{Di Matteo et al.}{2002}]{2002ApJ...564..576D} Di Matteo T., Perna R., Abel T., Rees M.~J., 2002, ApJ, 564, 576 

\bibitem[\protect\citeauthoryear{Fan et al.}{2006}]{2006AJ....132..117F} Fan X., et al., 2006, AJ, 132, 117 

\bibitem[\protect\citeauthoryear{Feroz, Hobson \& Bridges}{2009}]{2009MNRAS.398.1601F} Feroz F., Hobson M.~P., Bridges M., 2009, MNRAS, 398, 1601 

\bibitem[\protect\citeauthoryear{Field}{1959}]{1959ApJ...129..525F} Field G.~B., 1959, ApJ, 129, 525 

\bibitem[\protect\citeauthoryear{Field}{1958}]{1958PIRE...46..240F} Field G.~B., 1958, Proc. IRE, 46, 240 

\bibitem[\protect\citeauthoryear{Furlanetto \& Mesinger}{2009}]{2009MNRAS.394.1667F} Furlanetto S.~R., Mesinger A., 2009, MNRAS, 394, 1667 

\bibitem[\protect\citeauthoryear{Furlanetto, Oh, \& Briggs}{2006}]{2006PhR...433..181F} Furlanetto S.~R., Oh S.~P., Briggs F.~H., 2006, PhR, 433, 181 

\bibitem[\protect\citeauthoryear{Greig \& Mesinger}{2017}]{2017MNRAS.472.2651G} Greig B., Mesinger A., 2017, MNRAS, 472, 2651

\bibitem[\protect\citeauthoryear{Greig \& Mesinger}{2015}]{2015MNRAS.449.4246G} Greig B., Mesinger A., 2015, MNRAS, 449, 4246 

\bibitem[\protect\citeauthoryear{Greig et al.}{2017}]{2017MNRAS.466.4239G} Greig B., Mesinger A., Haiman Z., Simcoe R.~A., 2017, MNRAS, 466, 4239 

\bibitem[\protect\citeauthoryear{Hills et al.}{2018}]{2018Natur.564E..32H} Hills R., Kulkarni G., Meerburg P.~D., Puchwein E., 2018, Natur, 564, E32

\bibitem[\protect\citeauthoryear{Hoag, et al.}{2019}]{2019arXiv190109001H} Hoag A., et al., 2019, arXiv e-prints, arXiv:1901.09001

\bibitem[\protect\citeauthoryear{Hogg}{1999}]{1999astro.ph..5116H} Hogg D.~W., 1999, preprint (astro-ph/9905116)

\bibitem[\protect\citeauthoryear{Hurley-Walker et al.}{2017}]{2017MNRAS.464.1146H} Hurley-Walker N., et al., 2017, MNRAS, 464, 1146 

\bibitem[\protect\citeauthoryear{Jeli{\'c} et al.}{2008}]{2008MNRAS.389.1319J} Jeli{\'c} V., et al., 2008, MNRAS, 389, 1319 

\bibitem[\protect\citeauthoryear{La Porta et al.}{2008}]{2008A&A...479..641L} La Porta L., Burigana C., Reich W., Reich P., 2008, A\&A, 479, 641

\bibitem[\protect\citeauthoryear{Lane et al.}{2014}]{2014MNRAS.440..327L} Lane W.~M., Cotton W.~D., van Velzen S., Clarke T.~E., Kassim N.~E., Helmboldt J.~F., Lazio T.~J.~W., Cohen A.~S., 2014, MNRAS, 440, 327 

\bibitem[\protect\citeauthoryear{Liu \& Tegmark}{2011}]{2011PhRvD..83j3006L} Liu A., Tegmark M., 2011, Phys. Rev. D, 83, 103006 

\bibitem[\protect\citeauthoryear{Liu et al.}{2016}]{2016PhRvD..93d3013L} Liu A., Pritchard J.~R., Allison R., Parsons A.~R., Seljak U., Sherwin B.~D., 2016, PhRvD, 93, 043013 

\bibitem[\protect\citeauthoryear{Liu, Zhang, \& Parsons}{2016}]{2016ApJ...833..242L} Liu A., Zhang Y., Parsons A.~R., 2016, ApJ, 833, 242 

\bibitem[\protect\citeauthoryear{Longair}{2011}]{2011hea..book.....L} Longair M.~S., 2011, High Energy Astrophysics. Cambridge Univ. Press, Cambridge, UK  

\bibitem[\protect\citeauthoryear{Mao et al.}{2008}]{2008PhRvD..78b3529M} Mao Y., Tegmark M., McQuinn M., Zaldarriaga M., Zahn O., 2008, PhRvD, 78, 023529 

\bibitem[\protect\citeauthoryear{Mason et al.}{2018}]{2018ApJ...856....2M} Mason C.~A., Treu T., Dijkstra M., Mesinger A., Trenti M., Pentericci L., de Barros S., Vanzella E., 2018, ApJ, 856, 2 

\bibitem[\protect\citeauthoryear{McQuinn et al.}{2006}]{2006ApJ...653..815M} McQuinn M., Zahn O., Zaldarriaga M., Hernquist L., Furlanetto S.~R., 2006, ApJ, 653, 815 

\bibitem[\protect\citeauthoryear{Mellema et al.}{2013}]{2013ExA....36..235M} Mellema G. et al., 2013, Exp. Astron., 36, 235 

\bibitem[\protect\citeauthoryear{Mertens, Ghosh \& Koopmans}{2018}]{2018MNRAS.478.3640M} Mertens F.~G., Ghosh A., Koopmans L.~V.~E., 2018, MNRAS, 478, 3640

\bibitem[\protect\citeauthoryear{Mesinger \& Furlanetto}{2007}]{2007ApJ...669..663M} Mesinger A., Furlanetto S., 2007, ApJ, 669, 663 

\bibitem[\protect\citeauthoryear{Mesinger, Ewall-Wice \& Hewitt}{2014}]{2014MNRAS.439.3262M} Mesinger A., Ewall-Wice A., Hewitt J., 2014, MNRAS, 439, 3262 

\bibitem[\protect\citeauthoryear{Mesinger, Ferrara \& Spiegel}{2013}]{2013MNRAS.431..621M} Mesinger A., Ferrara A., Spiegel D.~S., 2013, MNRAS, 431, 621 

\bibitem[\protect\citeauthoryear{Mesinger, Furlanetto \& Cen}{2011}]{2011MNRAS.411..955M} Mesinger A., Furlanetto S., Cen R., 2011, MNRAS, 411, 955 

\bibitem[\protect\citeauthoryear{Morales \& Hewitt}{2004}]{2004ApJ...615....7M} Morales M.~F., Hewitt J., 2004, ApJ, 615, 7 

\bibitem[\protect\citeauthoryear{Morales \& Wyithe}{2010}]{2010ARA&A..48..127M} Morales M.~F., Wyithe J.~S.~B., 2010, ARA\&A, 48, 127 

\bibitem[\protect\citeauthoryear{Morales, Bowman \& Hewitt}{2006}]{2006ApJ...648..767M} Morales M.~F., Bowman J.~D., Hewitt J.~N., 2006, ApJ, 648, 767 

\bibitem[\protect\citeauthoryear{Mozdzen et al.}{2017}]{2017MNRAS.464.4995M} Mozdzen T.~J., Bowman J.~D., Monsalve R.~A., Rogers A.~E.~E., 2017, MNRAS, 464, 4995

\bibitem[\protect\citeauthoryear{Paciga et al.}{2013}]{2013MNRAS.433..639P} Paciga G. et al., 2013, MNRAS, 433, 639 

\bibitem[\protect\citeauthoryear{Parsons et al.}{2010}]{2010AJ....139.1468P} Parsons A.~R. et al., 2010, AJ, 139, 1468 

\bibitem[\protect\citeauthoryear{Parsons et al.}{2012}]{2012ApJ...756..165P} Parsons A.~R., Pober J.~C., Aguirre J.~E., Carilli C.~L., Jacobs D.~C., Moore D.~F., 2012, ApJ, 756, 165 

\bibitem[\protect\citeauthoryear{Planck Collaboration et al.}{2016}]{2016A&A...596A.108P} Planck Collaboration, et al., 2016, A\&A, 596, A108 

\bibitem[\protect\citeauthoryear{Planck Collaboration et al.}{2016}]{2016A&A...594A..13P} Planck Collaboration, et al., 2016, A\&A, 594, A13 

\bibitem[\protect\citeauthoryear{Pober et al.}{2013}]{2013ApJ...768L..36P} Pober J.~C. et al., 2013, ApJ, 768, L36 

\bibitem[\protect\citeauthoryear{Pritchard \& Loeb}{2012}]{2012RPPh...75h6901P} Pritchard J.~R., Loeb A., 2012, RPPh, 75, 086901 

\bibitem[\protect\citeauthoryear{Shaver et al.}{1999}]{1999A&A...345..380S} Shaver P.~A., Windhorst R.~A., Madau P., de Bruyn A.~G., 1999, A\&A, 345, 380 

\bibitem[\protect\citeauthoryear{Sims et al.}{2016}]{2016MNRAS.462.3069S} Sims P.~H., Lentati L., Alexander P., Carilli C.~L., 2016, MNRAS, 462, 3069 

\bibitem[\protect\citeauthoryear{Sims et al.}{2019}]{2019MNRAS.484.4152S} Sims P.~H., Lentati L., Pober J.~C., Carilli C., Hobson M.~P., Alexander P., Sutter P.~M., 2019, MNRAS, 484, 4152

\bibitem[\protect\citeauthoryear{Taylor, Carilli, \& Perley}{1999}]{1999ASPC..180.....T} Taylor G.~B., Carilli C.~L., Perley R.~A., 1999, ASPC, 180

\bibitem[\protect\citeauthoryear{Tingay et al.}{2013}]{2013PASA...30....7T} Tingay S.~J. et al., 2013, Publ. Astron. Soc. Aust., 30, 1 

\bibitem[\protect\citeauthoryear{van Haarlem et al.}{2013}]{2013A&A...556A...2V} van Haarlem M.~P. et al., 2013, A\&A, 556, A2 

\bibitem[\protect\citeauthoryear{Wayth et al.}{2015}]{2015PASA...32...25W} Wayth R.~B., et al., 2015, PASA, 32, e025 

\bibitem[\protect\citeauthoryear{Wilman et al.}{2008}]{2008MNRAS.388.1335W} Wilman R.~J. et al., 2008, MNRAS, 388, 1335 

\bibitem[\protect\citeauthoryear{Wouthuysen}{1952}]{1952AJ.....57R..31W} Wouthuysen S.~A., 1952, AJ, 57, 31 

\bibitem[\protect\citeauthoryear{Zheng et al.}{2012}]{2012MNRAS.424.2562Z} Zheng Q., Wu X.-P., Gu J.-H., Wang J., Xu H., 2012, MNRAS, 424, 2562 


\end{thebibliography}
\end{document}